# Methodological challenges of scenario generation validation: a rear-end crash-causation model for virtual safety assessment

Jonas Bärgman[1)], Malin Svärd[2)], Simon Lundell[2)] and Erik Hartelius[2)]
1) Division of Vehicle Safety, Department of Applied Mechanics, Chalmers University of Technology
2) Volvo Car Corporation, Volvo Cars Safety Centre

**Abstract**

Safety assessment of crash and conflict avoidance systems is important for both the automotive industry and other stakeholders. One type of system that needs such an assessment is a driver monitoring system (DMS) with some intervention (e.g., warning or nudging) when the driver looks off-road for too long. Although using computer simulation to assess safety systems is becoming increasingly common, it is not yet commonly used for systems that affect driver behavior, such as DMSs. Models that generate virtual crashes, taking crash-causation mechanisms into account, are needed to assess these systems. However, few such models exist, and those that do have not been thoroughly validated on real-world data. This study aims to address this research gap by validating a rear-end crash-causation model which is based on four crash-causation mechanisms related to driver behavior: a) off-road glances, b) too-short headway, c) not braking with the maximum deceleration possible, and d) sleepiness (not reacting before the crash). The pre-crash kinematics were obtained from the German GIDAS in-depth crash database. Challenges with the validation process were identified and addressed. Most notably, a process was developed to transform the generated crashes to mimic the crash severity distribution in GIDAS. This step was necessary because GIDAS does not include property-damage-only (PDO) crashes, while the generated crashes cover the full range of severities (including low-severity crashes, of which many are PDOs). Our results indicate that the proposed model is a reasonably good crash generator. We further demonstrated that the model is a valid method for assessing DMSs in virtual simulations; it shows the safety impact of shorter 'longest' off-road glances. As expected, 'cutting away' long off-road glances substantially reduces the number of crashes that occur and reduces the average delta-v.

This work highlights the need to both a) thoroughly understand the process of generating virtual scenarios and b) have the tools to validate them. While more work to develop validation processes for scenario generation is needed across all levels of crash severity, the transform and other validation tools that were developed bring us one step closer to accurate validation methodologies.

**Keywords:** crash-causation; scenario generation; validation; driver model; safety assessment; counterfactual simulations

# 1 Introduction

It is important that developers, as well as other stakeholders including governments and consumer rating organizations (e.g., NCAP; A. Miller et al., 2014), have the tools to efficiently and reliably assess the impact of safety systems. Safety systems include conflict avoidance systems (e.g., Adaptive Cruise Control: ACC and higher levels of automation such as Automated Driving Systems: ADSs) and crash avoidance systems (e.g., Automated Emergency Braking: AEB and other Advanced Driver Assistance Systems: ADASs).

However, it is not only the safety systems that need to be assessed with respect to traffic safety, but also factors affecting driver behavior (which in turn affect safety), such as different sources of driver distraction and measures taken to mitigate the impact of such factors. Driver distraction has been shown to have a substantial negative impact on safety (Klauer et al., 2014; Victor et al., 2015). As for measures to mitigate the safety impact of distraction, there is currently much focus on driver monitoring systems (DMSs), which aim to impact driver behavior in order to make driving safer. Part of the reason for this emphasis is European legislation (European_Parliament, 2019).

There are many ways to assess safety, but the methods available may be limited because of a lack of real-world data—for example, during the system's development. Note that here the term 'system' is used in a broad sense: it may be a conflict or crash avoidance system, an in-vehicle human machine interface (HMI; e.g., radio controls), a distraction task that is brought into the vehicle (e.g., a smartphone), or a measure to mitigate consequence of distraction and inattention (e.g., DMSs). One assessment method that has been receiving much attention recently is the use of virtual simulations to compare 'baseline' traffic situations (without the system) with 'treatment' traffic situations (with the system virtually applied). The comparison typically entails assessing the proportion of baseline crashes that are completely avoided in the treatment simulations, but the mean of the impact speeds or injury risk (or some metric describing their entire distributions) for the baseline and treatment simulations can also be compared. Note that using virtual simulations to assess conflict and crash avoidance systems is now common (Bjorvatn et al., 2021; Wimmer, 2023); however, using them to assess the behavior of a distracted driver and the systems that mitigate its consequences are still rare (with some exceptions, e.g., Bärgman, Boda, & Dozza, 2017; Bärgman, Lisovskaja, Victor, Flannagan, & Dozza, 2015; J. Y. Lee, Lee, Bärgman, Lee, & Reimer, 2018).

A core component of the virtual-simulation method is an accurate, realistic baseline—the set of conflict situations to which the system under assessment is applied for comparison. There are several different ways to create a baseline for a specific assessment. One way is to directly use digital representations of actual (reconstructed) crashes (Schubert, 2013). Another is to create them synthetically (Wimmer, 2023).

There are several reasons why synthetic baseline generation is more enticing than using actual crash data. Firstly, crashes are rare, which means that the statistical power of an assessment that only uses actual crashes is often limited. Secondly, collecting data from actual crashes is very costly. Creating synthetic crashes in a virtual environment is likely to be substantially cheaper. Thirdly, crash data represent the crashes of "yesterday"; they are not necessarily representative of the crashes of today or the future. If synthetically generated crashes can truly represent a real-world crash distribution, then they can ensure the validity of safety system assessments and improve the (virtual) prospective assessments of safety technologies in traffic environments of the future.

There are different ways to generate synthetic baselines: a) without simulation, by sampling from a multi-dimensional joint distribution space which describes the pre-crash kinematics in detail, either as variations of actual crashes (Leledakis et al., 2021) or of crashes created synthetically from distributions (Wimmer, 2023); b) through virtual simulations of traffic interactions based on models of road user interactions, where the initial conditions are based on distributions (Fries, Fahrenkrog, Donauer, Mai, & Raisch, 2022; Helmer, Wang, Kompass, & Kates, 2015; van Lint & Calvert, 2018); and c) through virtual simulations that generate variations of actual conflicts, by keeping some of the actual conflict kinematics while replacing others with driver behavior models (Bärgman et al., 2015; Bärgman & Victor, 2019; J. Y. Lee et al., 2018).

Whichever method is used to create synthetic crashes, the outcome should be a reasonable representation of the real-world scenario they seek to represent. If the baseline does not represent the real-world the impact assessment will likely not be accurate. The measures of interest—such as, for rear-end crashes, the distribution of impact speeds or the change in speed during the crash (delta-v) of the generated crashes—should be reasonably similar for the synthetically generated baseline crashes and the actual crashes they represent. This work set out to validate a method, based on crash-causation research, aimed to generate realistic baseline scenarios for rear-end crashes.

The behavior-based crash-causation model (hereafter called the CBM for convenience) is explicitly based on crash-causation research (Bärgman et al., 2017; Hautzinger, Pfeiffer, & Schmidt, 2005; Markkula, Engström, Lodin, Bärgman, & Victor, 2016; Victor et al., 2015). It consists of four sub-models, each named after the rear-end crash-causation mechanism that it describes: 1) Off-road glances (when off-road glances delay the braking response), 2) Too-close (when drivers are driving too close to the lead vehicle), 3) Low-deceleration (when drivers do not brake as hard as the car and the environment permit), and 4) No-response (when drivers are sleepy or drowsy).

As for the first mechanism, several studies have shown off-road glances to be a main contributing factor to crashes (Hickman, Hanowski, & Bocanegra, 2010; Klauer et al., 2014; Victor et al., 2015), especially in rear-end crashes. In fact, glances off-road have been shown to substantially increase crash and injury risks in manual driving (Klauer et al., 2014; Victor et al., 2015). Note that although several studies have investigated details of the relationship between driver glance behavior and crash-causation (J. D. Lee, McGehee, Brown, & Reyes, 2002; Markkula et al., 2016; Victor et al., 2015)—for example, as part of assessing DMSs, few have investigated the relationship between glance behavior and crash outcome (e.g., delta-v and crash avoidance). Further, many drivers follow a lead vehicle too closely, so crashes can occur even when drivers have their eyes on the forward roadway. This situation is described by Too-close, the second crash-causation mechanism.

Further, when drivers are about to crash, they often do not brake as hard as possible, in terms of what the car is capable of and what the environment (such as road surface) would support (Low-deceleration, the third crash-causation mechanism; Bärgman et al., 2017; Fajen, 2005; Kusano & Gabler, 2012). That is, they might have been able to avoid the crash, or at least lower the crash speed, if they had braked harder. This crash-causation mechanism is 'independent of' and 'in addition to' the first two mechanisms.

Finally, the fourth crash-causation mechanism described in one of the sub-models is No-response drivers. While Knipling and Wang (1994) estimate that the proportion of rear-end crash cases caused by sleepy drivers is 6.6%, others have estimated the proportion to be between 9% and 20% (Horne & Reyner, 1999). The No-response mechanism includes all following-vehicle drivers who are fatigued or sleepy, and therefore completely fail to brake in response to the evolving conflict with the lead vehicle, causing a crash at maximum impact speed (i.e., there is no reduction in speed of the following vehicle during the event).

Together, these four crash-causation mechanisms are included in the CBM model to be validated in this work because they account for a large proportion of rear-end crashes on our roads (Horne & Reyner, 1999; Kiefer, Salinger, & Ference, 2005; Klauer et al., 2014). However, the fact that the proportion of crashes caused by each individual crash-causation mechanism is not known with certainty may impact the validity of theory-based, baseline-scenario generation models, such as the CBM in this study.

With respect to the validation process of synthetically generated baselines, it is not obvious what to validate against. Optimally, data from real-world crash databases that include **all** levels of severity should be used (unfortunately, such data are typically not available). While crashes generated through simulation typically include all levels of severity (even the lowest-severity, property-damage-only (PDO) crashes), real-world databases do not. Specifically, if validation is to be performed comparing the outcome of synthetically generated baseline crashes with crashes from a crash database with a selection bias (in practice, all currently available crash databases), that bias must be considered in order for the comparison to be meaningful. Basically, all crash databases censor lower-severity crashes for failing to meet some inclusion criteria. For example, they might only include crashes when someone is injured (e.g., German In Depth Accident Study; GIDAS; Schubert, 2013), when repair costs are greater than a given threshold (Isaksson-Hellman & Norin, 2005; Ydenius, Stigson, Kullgren, & Sunnevång, 2013), or when a vehicle has to be towed from the scene (US-DOT, 1998). In any case, censored crashes will not be in the database. One may argue that the application of injury risk curves mitigates the bias effect, but we found no studies quantifying to what extent the bias is mitigated if this is done. Also, the authors of the current paper have not found any literature that specifically investigates other methods of mitigating the selection bias to enable accurate validation of the baseline generation.

Fortunately, when it comes to quantifying the selection bias itself, studies have estimated the proportion of PDO crashes in crashes of all severity levels (of a specific scenario type). The proportion is typically large. For example, Knipling et al. (1992) estimated that approximately 70-75% of all rear-end crashes are PDO crashes, and Blincoe et al. (2023) estimated that 73% of all crashes are PDOs (see Appendix A for details).

There are, however, databases that include some PDO crashes, such as the Folksam insurance company's database (Ydenius et al., 2013). For inclusion in their database, the crashes must incur repair costs above a certain threshold; less costly PDO crashes are consequently missing. Since there are estimates of the overall proportion of PDO crashes (at least for some scenarios), it should be possible to complement the PDO crashes that are available (e.g., in the Folksam database) with estimates of the ones that are missing.

This work set out to validate a model that generates baseline rear-end crash scenarios using four known crash-causation mechanisms. Consequently, a main objective of the work was to validate the proposed CBM against reconstructed crash data (specifically, GIDAS data). The validation included sensitivity analyses and a comparison of CBM's delta-v with GIDAS and that of a simple brake-light-onset-based crash-causation model (BLOM). However, limitations of the validation process emerged during the work, creating a second objective: to propose a method to account for the selection bias in the GIDAS delta-v data when used for validation of baseline generation. A tertiary objective was to compare the outcome delta-vs and crash avoidance between SHRP2 baseline glance behavior and the glance behaviors resulting from two hypothetical DMSs with interventions (to reduce long glances off-road). The DMSs are ideal, in the sense that they completely remove off-road glances longer than a specific duration.

The primary metric used to assess and compare models was delta-v, calculated based on the relative speed at impact, vehicle masses, and conservation of momentum (see Appendix B for a detailed description of how delta-v is estimated in his work). Thus visualizations of the delta-v distributions are the primary means of comparison, but delta-v mean and other distribution-related summary metrics are also used. In addition, the percentage avoided crashes is used to assess the ideal, hypothetical DMSs.

## 2 Method

### 2.1 Data and data pre-processing for scenario generation

This study used three different types of data to generate crashes: crash kinematics data, glance behavior data, and driver pre-crash deceleration data. The three types and the pre-processing performed for each are described below.

### 2.1.1 Crash kinematics data (from the GIDAS database)

These data were collected from road traffic crashes from two metropolitan areas in Germany, Dresden and Hanover (Otte, 2003; Schubert, 2013). For a crash to be included, at least one person must have been injured as a consequence of the crash. Each case is entered into the database with codes for all the information from the crash investigator's compilation (e.g., police report, photos, vehicle data, hospital information, and interviews or surveys of the precrash phase). GIDAS also offers a subset of cases in a time-series format, a pre-crash matrix (PCM). In these cases the following vehicle's trajectories and acceleration profiles for the five seconds leading up to the collision are reconstructed in 10ms steps. The PCM data used in this study were purchased by Volvo Car Corporation (VCC; which regularly purchases a subset of both GIDAS and PCM data). Only the trajectories and information (e.g., size and wheelbase) of the involved vehicles were used for the simulations. Delta-v information from the GIDAS reconstructions was used for delta-v calculation validation (Appendix B). All processing and handling of GIDAS data were done by VCC employees.

The preliminary criteria for selecting GIDAS cases in this study were: 1) PCM data is available and 2) the conflict situation corresponds to the "rear-end frontal" GIDAS crash type, defined by the Volvo Safety Center as a following vehicle crashing into a lead vehicle, when both are traveling in the same direction in the same lane. Data from the years 2008-2022 were selected and all crashes that fulfilled the above criteria were included.

The crashes were then further filtered using the following criteria: a) no lateral (steering) avoidance maneuver; b) no identified intention to overtake or change lanes; c) no on- or off-ramp nearby; d) the following-vehicle was not in a crossing; e) speed limit >50 kph; f) daylight, dusk, or dawn; and g) road was dry, damp, or wet.

All cases that fulfilled the above criteria were included, except for six cases which either had quality issues in the PCM or which were outside the range of technical limitations in the simulation methods, resulting in a total of 103 rear-end frontal crashes and were consequently used for this study.

### 2.1.2 Glance behavior data from SHRP2 and VCC (naturalistic driving studies)

Two sets of off-road glance-behavior data from baseline driving were used in this study: a) data from an internal VCC study (hereafter called the Kungälv study; see Bärgman & Victor, 2019 for details about the glance data) and b) data from the Second Strategic Highway Research Program (SHRP2) Naturalistic Driving Study (Blatt et al., 2015). Descriptions of these datasets follow. Note that off-road glance data from the SHRP2 study was used as the primary glance data source, while the Kungälv data was used for a sensitivity analysis (i.e., to check how the specific baseline glance distributions affect the delta-v of the generated baseline crashes).

The Kungälv study was conducted on a main highway in the Gothenburg area. Twenty participants (9 women and 11 men) aged 27 to 62 drove a Volvo MY2016 XC90 between Gothenburg and Kungälv, conducting a set of self-paced visual manual tasks during the drive. The participants, recruited through emails sent to Volvo employees, were required to meet certain criteria—such as having a minimum of 5000 kilometers of driving experience in the previous year and not being involved in vehicle development or working as test drivers. In the original study, visual manual tasks at different levels of automation were performed. However, the current study, designed and conducted afterwards, used only the baseline data collected during manual driving. That is, for each driver, a 40-second data segment, starting 30 seconds after the completion of the last task, was used as the baseline.

The glances from the Kungälv data were manually coded, frame-by-frame, as located either on- or off-road. Glance transitions and blinks were coded as part of the previous glance (whether on- or off-road). However, if the glance was moved from on-road to off-road during a blink, the blink was considered part of the following off-road glance instead. The duration of all off-road glances (for all drivers) were put into bins with a width of 0.1 s, while the total proportion of time the drivers looked on the roadway added as a point mass at the off-road glance duration of zero (i.e., a single point in the distribution). The data are shown in Figure 1.

The SHRP2 data were obtained from the SHRP2, the largest in-vehicle naturalistic driving study conducted to date (Blatt et al., 2015). During this study, 3247 primary participants drove instrumented vehicles in their everyday lives across six sites in the US, covering a distance of almost 80 million kilometers over a two-year period. The recordings captured not only everyday driving but also nearly 7000 near-crashes and over 1000 crashes (severity levels 1-3; see (VTTI, 2015). The present study used driver glance data manually annotated from a subset of everyday driving (i.e., baseline) data from the Victor et al. (2015) study. Specifically, the off-road glance behavior data of the baseline epochs (segments 30s long) were extracted to match the SHRP2 rear-end crashes and near-crashes used in the Victor et al. study. A total of 1791 epochs (segments of time), with a total of 4604 off-road glances, were obtained.

The glance behavior information from the SHRP2 data was extracted in the following way: The driver's glance direction at a given time step was given by a discrete signal, with each value corresponding to a specific location (e.g., "Forward", "Instrument Cluster", or "Cellphone"). The driver's glance was considered to be on-road when the glance location was coded as either "Forward" or "Left windshield", and off-road otherwise. Glance transitions and samples corresponding to eyes closed, as well as short sequences (< 1s) of invalid values were concatenated to the previous glance (whether it being on- or off-road). Long sequences (> 1s) of invalid values were instead substituted with a single on-road glance sample, to avoid introducing too much uncertainty in the glance location data and to ensure separation of off-road glances. The durations of all off-road glance sequences were collected and sorted into bins of 0.1s each. The data are shown in Figure 1.

The percents of on-road are slightly different for SHRP2 and Kungälv. Both are, however, close to 80% at zero seconds eyes-off road (i.e., eyes-on-road), which is a typical value for baseline driving (Victor, 2005).

The cumulative distribution looks relatively similar (left panel; Figure 1), but in the probability density function (right panel) it is clear that Kungälv has some more off-road glances between 1 and 2 s, while it does not have any glances beyond 2.6s (whereas the longest glance for SHRP2 is 6.7s).

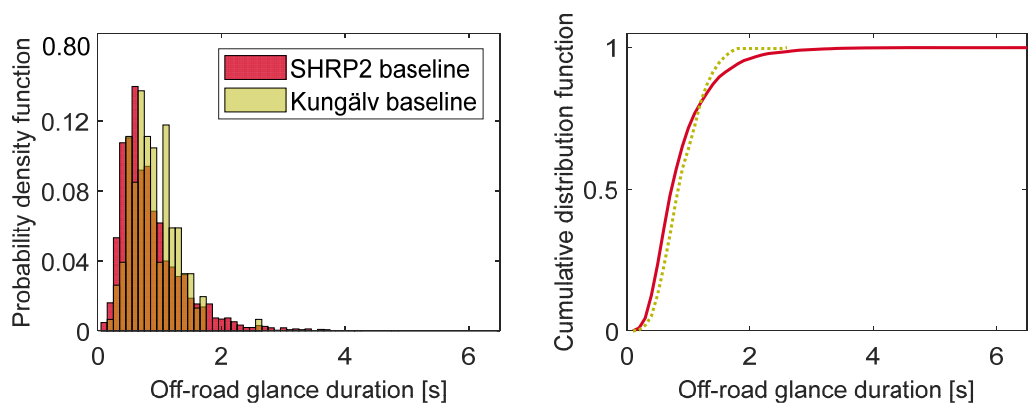

*Figure 1. The empirical probability density function of the off-road glance durations (left), and the cumulative density function of the same data, ignoring the point mass at zero (it would start at that value on the y-axis). The on-road glances (point-mass at zero) were not included here, as they are very similar between the two distributions.*

### 2.1.3 SHRP2 glances with hypothetical driver monitoring systems

To illustrate how counterfactual simulations using the CBM model can be used to assess a driver monitoring system (DMS), we have created two ideal, hypothetical DMSs. In order to affect driver glance behavior, a DMS must provide the driver with some feedback (e.g., warning or nudging) when the driver looks away from the road for too long. Thus the two

DMSs both include driver feedback. As we are only demonstrating the method, we made the impact of the intervention on glances simple: we assumed that all glances beyond a time period of 3s (in the first DMS) or 2s (in the second DMS) are completely removed from the SHRP2 distribution. That is, these are ideal DMSs, with a glance-removal performance much better than any real-world system's: no driver will ever look away longer than the time period defined for that DMS. The shape of the remaining distribution, normalized to a probability density function, was retained. We assumed that drivers have the same proportion of on-road glances for the two DMSs – that of the original glance distribution (~80%). Figure 2 shows a comparison between the SHRP2 baseline and the 'cut versions', hereafter called SHRP2cut3s and SHRP2cut2s.

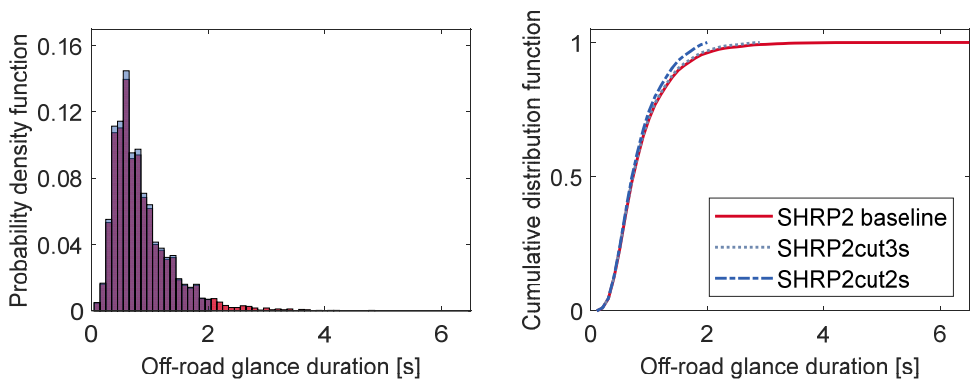

*Figure 2. The SHRP2 baseline off-road glances and the same but cut at 3s and 2s. as empirical probability (left panel) and cumulative (right panel) probability functions. As the point-mass is the same across all three, it is excluded. Note that SHRP2cut3s is not visible in the left panel as there is very little change compared to the original SHRP2 distribution.*

**2.1.4 Driver pre-crash deceleration data from SHRP2 (naturalistic driving study)**

While the onset of braking is modeled as in previous studies (Bärgman et al., 2017; Bärgman et al., 2015; Bärgman & Victor, 2019), the pre-crash braking behavior (i.e., not when but how hard the driver brakes) is described by a (more or less) fixed jerk and a distribution of maximum deceleration from real crashes.

In the simulation environment (using esmini; Knabe, 2023) as basis, with an external controller containing the driver, vehicle dynamics, sensors and the ADAS), the deceleration was controlled by applying a brake pedal force. As the method required the target deceleration to be available, the relation between deceleration and brake pedal force was identified empirically and mapped in the simulation software. In the simulation environment the jerk was controlled by applying a ramp-up of the force with which the driver pressed the brake pedal in the simulation environment. A mean jerk of -23.04 m/s³ was determined empirically. There were some fluctuations in jerk across the deceleration range, resulting in a standard deviation of 0.74 m/s³ over the desired deceleration range. The same jerk was applied irrespective of the targeted deceleration.

The maximum deceleration was described by an empirical distribution extracted from 45 rear-end crashes in the SHRP2 data (see Figure 3). The dataset came from the Victor et al. (2015)

study and the maximum deceleration was extracted from each case by fitting a piecewise linear model to the longitudinal acceleration data from each crash (see Figure 1 and Figure 2 in Markkula et al., 2016, for examples). Only crashes with a maximum braking plateau were included, to ensure they genuinely contain the drivers' maximum deceleration. That is, crashes in which the deceleration was still increasing at the time of the crash were excluded.

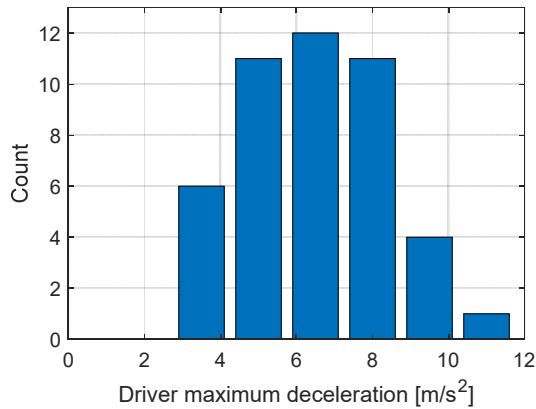

*Figure 3. The empirical maximum driver deceleration from 45 SHRP2 crashes used in this work*

### 2.2 The CBM's four sub-models

The CBM that this study aims to validate consists of four sub-models. The first, *Off-road glance*, is based on a response delay to a lead-vehicle conflict due to the following vehicle driver's off-road glances. The second, *Too-close*, is based on drivers who are driving too closely to the lead vehicle. The third, *Low-deceleration*, reflects a driver's failure to brake at the maximum deceleration possible (given the vehicle and road conditions). The fourth, *No-response*, relates to sleepy drivers. These sub-models are described in turn below. Note that the first sub-model, in addition to describing the crash-causation model, also includes a description of the model of the driver brake onset response, while the third sub-model includes a description of braking control (how the driver brakes).

#### 2.2.1 Sub-model 1: Off-road glance behavior-based crash-causation and response onset model

This sub-model quantifies how drivers' glances off-road limit their ability to respond to a looming lead vehicle in time to avoid a crash. The off-road glance itself creates a response delay, based on the fact that the driver of the following vehicle is not looking at the lead vehicle, which in this scenario is on the road in front. When the driver does look back at the road, there is a further perception-action delay. This delay includes detecting the looming lead vehicle, moving the foot from the accelerator pedal to the brake pedal, depressing the brake pedal, and any subsequent delay due to the mechanics of the vehicle (e.g., brake pressure buildup and brake pad movement). See Figure 4 for an illustration of the glance application process.

The sub-model assumes that drivers do not accumulate information about the lead vehicle when they are looking away from the forward roadway. Note that this is a simplification, as studies have shown that the eccentricity of the off-road glance (i.e., how great the angle is that represents the difference between looking at the road ahead and the driver's actual glance direction) affects the time it takes for the driver to respond after looking back at the road (Svärd, Bärgman, & Victor, 2021).

The specifics of the sub-model are based on the work presented in Markkula et al. (2016), which shows that drivers who look away from the roadway but look back again after $\tau^{-1}$=0.2s$^{-1}$ respond very fast, with an average response time of just below 0.5s. $\tau^{-1}$ is defined using the angle that the width of the lead vehicle subsumes on the driver's retina ($\theta$) and its derivative ($\dot{\theta}$) as $\tau^{-1} = \frac{\dot{\theta}}{\theta}$. It is the inverse of the optically defined time to collision (from the following-vehicle's driver's perspective) and is a measure of conflict urgency. The 0.2s$^{-1}$ threshold is based on Markkula et al. (2016) (see Figure 4 in that paper). The response time is slightly dependent on the urgency of the situation, with an average of 0.42s—but a couple of hundred milliseconds slower in less urgent events (close to $\tau^{-1}$=0.2s$^{-1}$ when the driver looks back)—and a couple of hundred milliseconds faster at more urgent events (larger $\tau^{-1}$). However, in this study we used a fixed response delay of 0.5s to denote the time it takes for the deceleration to start after the driver looks back at the road.

The Markkula et al. (2016) study also shows that if a driver looks back on the road before $\tau^{-1}$=0.2s$^{-1}$, there is no correlation between urgency and response time. One interpretation of these two findings—an interpretation that we use in our CBM—is that only glances that overlap $\tau^{-1}$=0.2s$^{-1}$ are relevant for crash-causation. If the drivers look back earlier, they will respond no later than $\tau^{-1}$=0.2s$^{-1}$, while if they look back at $\tau^{-1}$>0.2s$^{-1}$, they will respond immediately (after the 0.5s response delay). After $\tau^{-1}$=0.2$^{-1}$, drivers will not look away, as the event is too critical (and they will have already initiated braking).

This interpretation also means that only the portion of the off-road glance that extends beyond $\tau^{-1}$=0.2s$^{-1}$ is relevant for crash-causation (see Appendix C for more details). Consequently, to investigate the impact that different off-road glance behavior 'counterfactually' would have on a specific crash, it is possible to virtually apply a specific distribution of off-road glance durations to that crash (see Figure 1 for an example of an off-road glance distribution for everyday highway driving). This is done by overlapping individual off-road glance durations with the time in the crash kinematics where $\tau^{-1}$=0.2s$^{-1}$, and then running simulations using the response model described above. It is possible to perform a statistical transformation of the original off-road glance distribution to what is called an overshot distribution, and subsequently place (anchor) an overshot glance from the overshot distribution at $\tau^{-1}$=0.2s$^{-1}$. See Appendix C for a description of the overshot distribution. In this study an overshot distribution was used to reduce the number of simulations needed by a factor of approximately 30. Finally, note that before the glance behavior is added to the pre-crash kinematics, the speed of the evasive maneuver is removed from the following vehicle's speed profile (see Figure 4), to allow for model-based counterfactual behaviors.

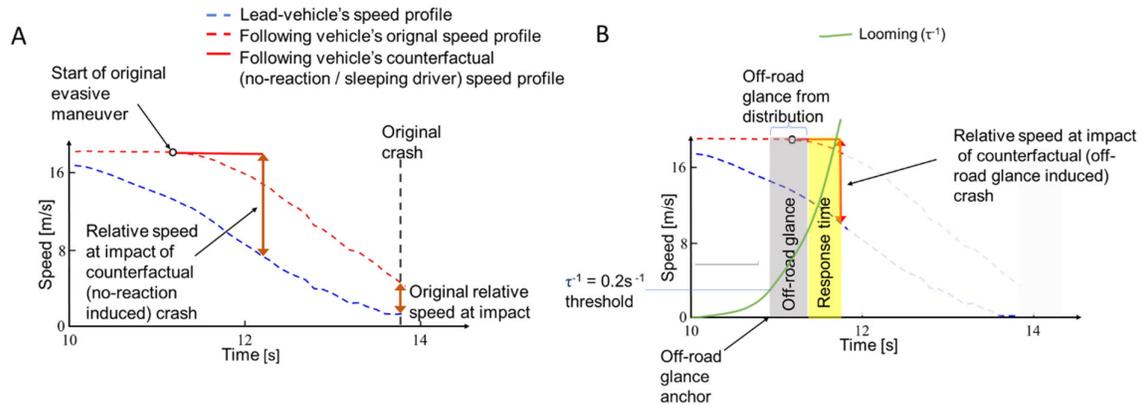

*Figure 4: Overall application figure, with looming, threshold, glance application, delay/response time. Rear-end crash with original speeds and no-reaction counterfactual speeds ("what-if" the following vehicle did not react; left panel), and an off-road glance induced counterfactual crash ("what-if" the driver looked away; right panel) with a glance anchor at $\tau^{-1}=0.2s^{-1}$.*

### 2.2.2 Sub-model 2: An attentive driver driving too close to the lead vehicle

The second crash-causation sub-model was implemented simply by having the driver look on-road at all times. This is operationalized by including the point mass at zero in the off-road glance distribution (see Figure 1) when performing the simulations. The crashes then occur due to the combination of short time gaps between the lead and following vehicles in the original crash kinematics and the maximum deceleration of the driver in each individual simulation (see Simulations below for details).

### 2.2.3 Sub-model 3: Drivers not braking as hard as possible

The third crash-causation sub-model reflects drivers' tendency to not brake as hard as possible, even if they are about to crash. The model's implementation is based on the distribution of drivers' maximum decelerations as described earlier. A previous study did not find any relationship between off-road glance duration and maximum deceleration in the SHRP2 data (Bärgman et al., 2017). Therefore, sub-models 1 and 3 are assumed to be independent.

### 2.2.4 Sub-model 4: Sleeping or drowsy drivers

The fourth and final crash-causation sub-model describes a sleepy driver who does not respond to the looming lead vehicle at all. It was assumed that this mechanism contributes to approximately 10% of all rear-end crashes (somewhere in the middle of what is reported in Horne & Reyner, 1999; Knipling & Wang, 1994). This sub-model was operationalized in the simulations as no braking at all prior to crashing. Practically the delta-v distribution of the no-response crashes (one per original GIDAS crash) was added so that the total proportion of these crashes out of all simulated crashes was 10%. That is, the no-response crashes were added after the crashes of the other three sub-models were created. The no-response crashes were created by simulating each event without any reaction by the following-vehicle driver. Practically, the final delta-v distribution was obtained by calculating the weighted sum of the (normalized) delta-v distribution resulting from the three previous sub-models (90%) and the (normalized) no-response delta-v distribution (10%).

## 2.3 The brake-light onset model (BLOM)

In addition to validating the CBM, we also compare the CBM and a simpler crash-causation model based on reaction times and brake-light onset, without considering glance behavior. In this model, known hereafter as BLOM, a driver starts braking (with some reaction time) after the lead-vehicle's brake lights go on. Note that in the literature crash generation causes crashes in various ways, but reaction time is almost aways one component. In fact, the reaction time to brake-light onset is very often used as a main safety metric in studies of human responses to critical situations (Aust, Engström, & Viström, 2013; Engström, 2010; Markkula, Benderius, Wolff, & Wahde, 2013). Also note that the BLOM **cannot** be used to assess visual distraction, as it does not include glance behavior.

The BLOM uses a simple driver reaction-time distribution and the same deceleration distribution as the CBM (see Figure 3). For each simulation a unique pair of reaction time ($T_{react}$) and maximum deceleration ($d_{max}$) from the distributions is used. In the simulation, the BLOM identifies the time point of the lead-vehicle start braking (the brake-light onset). It then waits $T_{react}$ seconds, after which it starts braking with $d_{max}$ (with the same variance on jerk and maximum deceleration as for the CBM; see 2.1.4).

We used the onset of the lead-vehicle braking as the reaction time onset, which is what most experiments measure when studying the impact on safety of a particular manipulation (e.g., when the driver is asked to perform a visual-manual or cognitive task (He & Donmez, 2022), or when assessing forward collision warning (Aust et al., 2013). However, using brake-light onset has its drawbacks, as for cases when the lead-vehicle is not braking at all or is standing still at the start of the event, the system would, depending on implemented logic, either not brake at all or brake fully directly at the start of the simulation, respectively. In this study we chose to exclude all cases that had these drawbacks to not unjustifiably classify a case as crash or avoidance when assessing BLOM, leading to a reduced dataset of 68 cases (35 excluded). This study used a log-normal reaction time distribution from Green (2000), with

$$\mu = \log\left(\frac{m^2}{\sqrt{v + m^2}}\right)$$

and

$$\sigma = \sqrt{\log\left(\frac{v}{m^2} + 1\right)}$$

where m= 1.275 and v = $0.6^2$. This distribution was for the purpose of simulations discretized into 25 bins, with a step size of 0.2s (from 0.2 to 5s), as shown in Figure 5. The distribution was cut at t=5s as the additional contribution of crashes beyond that is marginal.

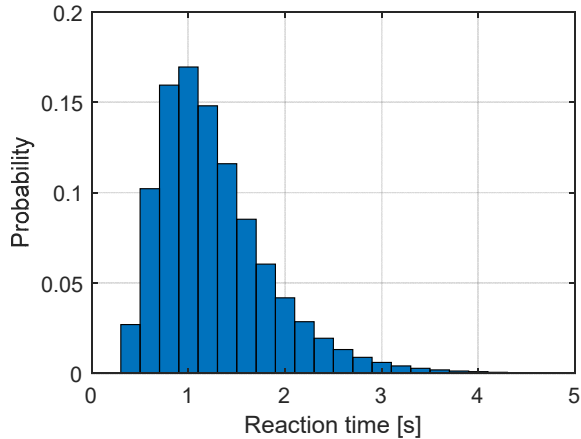

*Figure 5: The discretized log-normal reaction time distribution used in the BLOM.*

### 2.4 Parameter selection, simulations, and simulation output

The simulations were based on the 103 original rear-end GIDAS cases. The evasive maneuver of the following vehicle was removed from each case, replaced with a constant speed. The first step for each individual simulation was to select parameters to be used in the simulation, including setting the off-road glance duration (from the overshot-transformed baseline off-road glance distribution; see Figure 1) and the maximum deceleration (from the distribution shown in Figure 3).

### 2.5.1 Parameter selection

The off-road glance distribution was divided into 0.1s bins, creating a total of 27 bins for the Kungälv data and 67 bins for the SHRP2 data (as the longest glance is longer in the latter). The maximum deceleration was divided into 1.5 m/s$^2$ bins, creating a total of six bins. Nominally, each original crash with the SHRP2 glance data resulted in 67 x 6 = 402 simulations. With a total of 103 original crashes, the number of simulations in each simulation set was 103 x 402 = 41,406. A process to avoid running the same simulation more than once was adopted, substantially reducing the total number of simulations needed. See Appendix D for details.

### 2.5.2 Simulation sets

Seven sets of simulations were completed for this study. Table 1 shows the details. The 'Kungälv baseline' refers to the glance distribution collected for manual driving in Sweden, and the 'SHRP2 baseline' refers to the glance and deceleration distributions based on data from the SHRP2 project. 'Removed evasive maneuver' indicates whether the original evasive maneuver from the original conflict situation was preserved in the simulation. 'Reaction time' is for CBM the time from when the drivers had their eyes back on the road (after looking off-road) until they started braking (brake onset), and for BLOM the reaction time after the lead-vehicle brake light onset.

| Data/ model | Simulation type | Glance distribution | Deceleration distribution | Removed evasive maneuver | Looming threshold [$1/\tau$] | Reaction time [s] | No. simulations** Theoretical / Simulated / Crashes |
|---|---|---|---|---|---|---|---|
| GIDAS | Baseline | N/A | N/A | No | N/A | N/A | 103 / 103 / 103 |
| CBM | Baseline | SHRP2 baseline | SHRP2 baseline | Yes | 0.2 | 0.5 | 41406 / 16278 / 9951 |
| CBM | Baseline | Kungälv baseline | SHRP2 baseline | Yes | 0.2 | 0.5 | 16686 / 12261 / 4059 |
| BLOM | Baseline | N/A | SHRP2 baseline | Yes | N/A | distribution from Green* | 10200 / 5941 / 3088 |
| CBM | Baseline | SHRP2 baseline | SHRP2 baseline | Yes | 0.3 | 0.5 | 41406 / 14797 / 8997 |
| CBM | Baseline | SHRP2 baseline | SHRP2 baseline | Yes | 0.2 | 0.8 | 41406 / 20905 / 9312 |
| CBM | Driver Monitoring System; 1st DMS (ideal, hypothetical) | SHRP2 baseline cut at 3s | SHRP2 baseline | Yes | 0.2 | 0.5 | 18540 / 17441 / 4801 |
| CBM | Driver Monitoring System; 2nd DMS (ideal, hypothetical) | SHRP2 baseline cut at 2s | SHRP2 baseline | Yes | 0.2 | 0.5 | 12978 / 13127 / 2953 |

* Green here means the reaction time from Green (2000), also described in Section 2.3.

** 'Theoretical' means all possible permutations (all crashes and the combination of all bins in the off-road glance and deceleration distributions), 'simulated' means actual simulations executed (with logic to reduce the number of simulations required; see Appendix D), and 'crashes' means the number of unique crashes found (without prevalence weighting).

*Table 1: The simulation sets.*

### 2.5.3 Calculation of delta-v

We did not have enough information for the simulations to calculate the true change in speed during the crash of the involved vehicles (the typical definition of delta-v; see Wang, 2022). Instead, we used a process to convert the relative speed at impact, using the masses of the involved vehicles and assumptions about conservation of momentum. The details of how this delta-v was calculated for each simulated crash are provided in Appendix B, and further description about the validity of this calculation compared to GIDAS estimates of delta-v can be found in Appendix E.

### 2.6 Prevalence weighting of simulation crashes

When performing simulation-based virtual safety assessment, care needs to be taken to make the crashes resulting from simulations represent the original crashes. Specifically, each of the 103 original (seed; as they are the origins of the created crashes) GIDAS crashes typically generated a set of simulated crashes (see Figure 6 and Appendix J). Since each simulation crash with the CBM added can generate a different number of counterfactual crashes, the generated crashes must be prevalence-weighted so that the simulated crashes for a particular seed crash represent only the individual seed crash. That is, an original crash that results in ten simulated crashes should contribute to the overall outcome impact speed distribution to the same extent as an original crash that results in 100 simulated crashes. Appropriate prevalence weighting means that the contribution of each of the ten simulated crashes contributes 1/10, and each of the 100 simulated crashes contributes 1/100. Note that this balancing should be "on top" of accounting for the glance- and braking probabilities. The weighting was consequently accomplished as follows:

a. For each original crash, sum the probabilities of the combination of the glance and deceleration across all the simulations that ended up as crashes. Denote these sums as $q_1, ..., q_N$, where N is the number of original crashes.
b. Normalise the probabilities so that $q_1 + ... + q_N = 1$. $w_i = 1 / q_i$.
c. Re-weight the generated crashes by a factor $w_i$: for each original crash i and generated crash scenario j, use the weight $w_i\, p_j$, where $p_i$ is the product of the probability of the glance and deceleration of simulation j. Note that the sum of $w_i\, p_j$ at the end should be normalized to one.

Note that to ensure that no individual seed crash gets weighted unreasonably (which would result in that crash having too much impact on, for example, the delta-v distributions) in Step c, we used weight trimming (B. K. Lee, Lessler, & Stuart, 2011). Specifically, we limited the weights to those of the 5th and 95th percentile. That is, any seed crash that got a weight higher than the 95th percentile weight was set to the 95th percentile weight, and any seed crash with a weight lower than the 5th percentile weight was set to the 5th percentile weight.

The figures in Appendix F show the resulting simulated crashes per seed crash before and after prevalence weighting for CBM and BLOM. Those figures also illustrate how many of the simulated crashes crash at maximum delta-v (as if the driver did not react at all), before and after prevalence weighting.

## 2.7 Preparing for the comparison between GIDAS original and simulated data

A core part of validating a CBM such as the one proposed in this paper is to check whether the two datasets have the same underlying sample selection mechanisms—and if they do not, to ensure that any selection bias is compensated for.

This section describes a process to improve the validity of comparisons between simulated and actual crashes, when the latter are obtained from crash databases with selection bias (as in this case, since PDO crashes are not included and there is an underreporting of low severity crashes). Note that the process includes many assumptions, which are likely to impact the overall validation relevance; accordingly, they are listed below. We are, however, demonstrating the importance of performing this correction in order to validate crash-causation models.

First, we adjusted for the selection probability (e.g., underreporting) of low-severity crashes in GIDAS using data from Hautzinger et al. (2005). This was done by weighting on two levels of injury severity. As there were relatively few high-severity crashes (18 for CBM and 14 for BLOM), this adjustment only shifted the mean delta-v by approximately 0.7km/h towards lower delta-vs. Second, we sought to represent all rear-end PDO crashes by creating an analytical PDO distribution. We used insurance data with some (albeit censored) PDO data, and complemented the resulting distribution with an assumption about its shape as well as information from the literature. Using this distribution, we then added PDO data to GIDAS data, to achieve a GIDAS-with-PDO distribution. Finally, we estimated the parameters for an analytical transfer function that converts the GIDAS-with-PDO distribution to the original GIDAS data. We later also used this function to transform the delta-v distributions generated from simulations (which should include crashes on all levels of severity), in order to have the same sample selection as GIDAS. Details of this process are given below.

### 2.7.1 The Folksam data

The Folksam data come from insurance claims from vehicles with event data recorders (EDRs) which were installed by the insurance company for research purposes. The data has an inclusion threshold of 50,000 SEK (before 2012) and 70,000 SEK (from 2012) in repair costs (i.e., only crashes with repair costs above this are included in the data used in this study).

The data available for this study were at the occupant level only (not at the crash or vehicle level) and included MAIS levels 0-6 for each individual involved in the crashes for which data were collected. For each individual, only the maximum MAIS was available. For frontal impacts, only data for the driver and any front seat passenger were available. (See Ydenius et al., 2013, for more details about the data). A total of 912 data points (individuals) with MAIS0-6 coding and the corresponding estimated delta-v were available; whether the individual was a driver or a passenger was also specified.

In total, 43% of the individuals in the dataset had no injury (MAIS0); the remaining had injuries ranging from MAIS1-6. See Panel A in Figure 6 for the distribution of all crashes, and panel C for the distribution of MAIS0 individuals.

### 2.7.2 Data on the overall proportion of rear-end crashes with property damage only (PDO)

Studies have estimated the proportion of PDO crashes out of all crashes across all severities. One of the most cited sources on rear-end crashes is Knipling et al. (1992), which estimates that between 70% and 75% of all rear-end crashes incur property damage only (see Appendix A for details). A more recent study indicate 73% PDOs overall for US crashes (Blincoe et al., 2023). We use 70% in the PDO correction process in this work. However, a sensitivity analysis was performed to assess the impact of the choice of the overall proportion of PDOs in the data (Appendix G). The range 55% to 85% (i.e., 70% +-15%) with a step size of 5% was assessed.

### 2.7.3 Estimating the data censored by the repair-cost threshold

The intention was to create a complete PDO delta-v distribution, as accurate as the information currently available allows. Given the estimates of the total proportion of rear-end PDO crashes (70%) and of the proportion and delta-v distribution for the Folksam data (43%), we calculated how much data need to be added to the Folksam data to create a dataset representing all PDO cases, and where in the distribution they should be added.

- Step 1. Stating assumptions:
    1. Assume that repair cost increases as delta-v increases*.
    2. Similarly, assume that, for rear-end crashes in general, there are fewer crashes as delta-v increases*.
    3. From these assumptions it follows that the PDO crashes should be added to the Folksam data at the lowest delta-vs. It is also likely that the vast majority of the PDO data should be added below the mode of the Folksam PDO data (see Panel C in Figure 6).
    4. Let $P_{PDO}$=0.7
    5. Let $N_{Folksam\_PDO}$=0.43
- Step 2: Creating the PDO distribution:
    1. Identify the mode of the original Folksam PDO.
    2. Calculate how much data should be added to the Folksam PDO data to create a full PDO distribution.
        i. The total amount of PDO should be
            $N_{PDO\_total}$ = ($N_{Folksam\_MAIS0-6}$ - $N_{Folksam\_PDO}$) / (1 – $P_{PDO}$)
        ii. The amount of PDO that should be added should be
            $N_{Folksam\_PDO\_to\_add}$ = $N_{PDO\_total}$ - $N_{Folksam\_PDO}$
    3. Choose an analytical PDF form to be fitted to the full PDO data. Here we chose $f(\Delta v) = B_1/e^{(B_2 * \Delta v)}$.
    4. Manually add $N_{Folksam\_PDO\_to\_add}$ in the lower bins of the original Folksam PDO so that it "looks like" the chosen functional form. The data is consequently only added to the six lowest bins. **
    5. Find the parameters $B_1$ and $B_2$ that give the best fit.
    6. If the manually added $N_{Folksam\_PDO\_to\_add}$ does not create a good fit, return to Step 4., modifying the proportion across the lower bins until a satisfactory fit is obtained in Step 5.

*Note that we could not find specific support in the literature for Assumption 1 in Step 1. We did, however, talk to a researcher at a car insurance company, who confirmed that it is a reasonable assumption; it should be uncontroversial. Assumption 2 was also hard to confirm in the literature, as information about crashes with extremely low speeds/delta-vs are typically not captured in databases.*

*\*\* This is a simplification: there may also be data missing above the sixth bin in panel C (approximately 10km/h delta-v), and there may not be as much missing PDO at such "high" delta-vs. However, we do not have enough information to support a different approach. We chose the sixth bin based on the assumption of an exponential fit to the data (the data seems to drop off for lower delta-vs, and we did not want to go too far up). See also the sensitivity analysis in Appendix G.*

The output (Panel E in Figure 6) is consequently a best guess of the shape of the true PDO. Note that it does not have to be perfect—it is the approximate shape of the PDO distribution that is of interest, so that the GIDAS data is transformed to include PDO crashes. To assess the impact of how the data were added to the six lowest bins, we performed a sensitivity analysis. The results of this analysis can be found in Appendix G.

The final equation describing the full PDO distribution is:

$f(\Delta v) = 0.137/e^{(0.27 * \Delta v)}$ (Eq. 1)

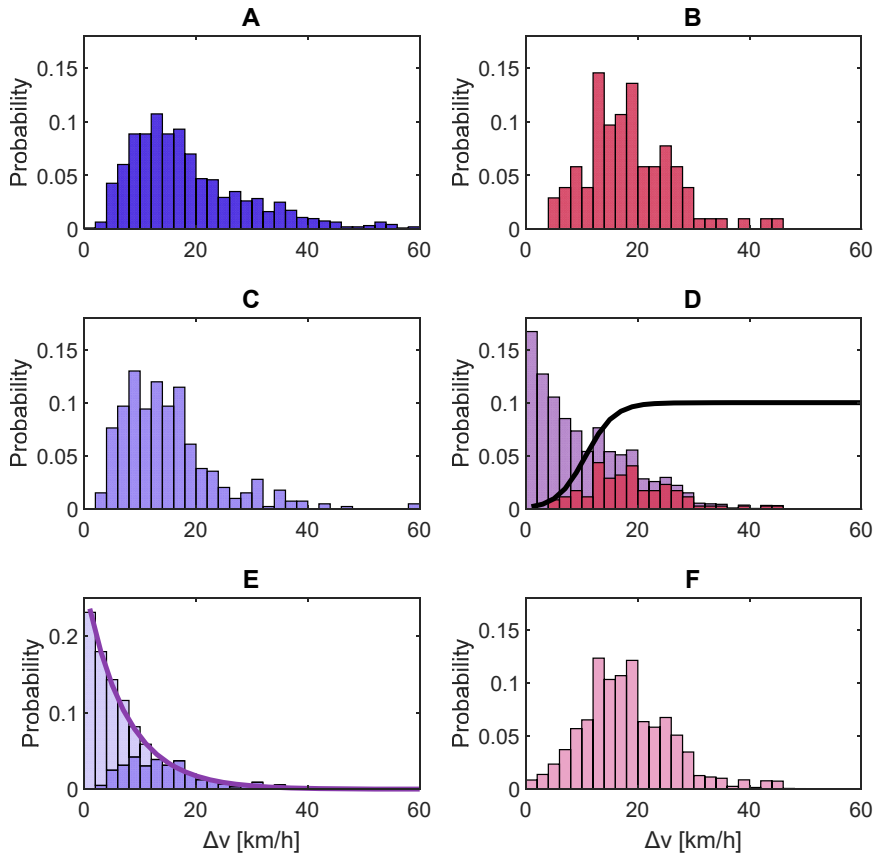

*Figure 6: The steps included in the creation of a transfer function from rear-end delta-v distributions that include all levels of severity to a delta-v distribution censored in the same way as GIDAS: Folksam data with 43% property damage only (PDO) crashes (Panel A). Folksam PDO data only (Panel C). Folksam PDO with data manually added at the lowest four bins, so that the total becomes 70% (complementing the 43%) to create an exponential decline and a fitted exponential (Panel E; Eq. 1; solid line). Original GIDAS crashes (Panel B). Original GIDAS crashes with data added per the fitted exponential in Panel E to make the total PDO 70%, including a transfer function (Eq. 2) to transform the data in Panel D to mimic the GIDAS sample selection. Panel F shows the outcome of the*
Page 18

*application of the transfer function (black line in Panel D) to the histogram in Panel D. Note that in panels E and D the full distribution is normalized to one; consequently, the original distributions are only illustrating their respective proportions of the full distribution (of which the original distribution is a part).*

### 2.7.4 Adding the PDO data to the GIDAS data

As the GIDAS data (ostensibly) only include injury data, we can add the PDO data (with the shape of the final equation in the last step) to the GIDAS data, with a proportion of PDO crashes from the literature (e.g., $P_{PDO}$ = 0.7). That is, we add data per Eq. 1 (see panel E in Figure 6) to the GIDAS original data (panel B in Figure 6), so that the proportion of PDO crashes is 70% of the total. The result can be seen in panel D in Figure 6.

### 2.7.5 Transforming the GIDAS data with PDO back to the original GIDAS data

The inclusion mechanism of GIDAS should now be the transform from the GIDAS with PDO (panel D in Figure 6) to the original GIDAS (without PDO; panel B in Figure 6). Such a transformation can be done in several ways, but here we chose to find $C_1$ and $C_2$ in

$$P_{MAIS0-6\_to\_MAIS1-6}(\Delta v) = e^{(C_1+C_2*\Delta v)} / (1 + e^{(C_1+C_2*\Delta v)})$$

by minimizing the cost function

$$\text{cost} = \text{sum}(\text{abs}(\text{GIDAS}_{original\_i} - \text{GIDAS}_{withPDO\_i} \cdot P_{MAIS0-6\_to\_MAIS1-6}(\Delta v_i)))$$

The cost is the sum of the absolute difference of all the bins in the histogram. That is, GIDAS with PDOs added (D in Figure 6) with the transform applied, subtracted from GIDAS original (B in Figure 6). The minimization of the cost was done by an exhaustive grid search of $C_1$ and $C_2$ (with the ranges -10 to -0.1 in 0.05 steps and from 0.001 to 5 in 0.001 steps, respectively) to find the pair of $C_1$ and $C_2$ with the lowest cost.

The resulting transfer function is

$$P_{MAIS0-6\_to\_MAIS1-6}(\Delta v) = e^{(-4.15 + 0.388*\Delta v)} / (1 + e^{(-4.15 + 0.388*\Delta v)}) \qquad \text{(Eq. 2)}$$

Note that any transformed distribution should either be normalized (to probability) or re-weighted (to the original number of cases). Equation 2 is shown as a sigmoid in Figure 6, Panel D. The application of that function to the distribution in Panel D is shown in Panel F. The distributions in Panel B and Panel F should now be similar, which they seem to be.
It should now be possible to use $P_{MAIS0-6\_to\_MAIS1-6}(\Delta v)$ directly to convert distributions created using crash-causation models that produce data across all severities (including MAIS0/PDOs), to mimic the GIDAS sample selection process.
### 2.8 Analysis

### 2.8.1 Comparison across GIDAS and models

Most comparisons are made by visualizing the complete empirical distribution of the delta-v of two simulations sets. The mean of the distribution is typically included in the figures. In Appendix H we also present statistics for some of the distribution comparisons and provide some examples of such metrics in the main manuscript.

Note that for most of the metrics several weighting steps are needed (i.e., case/propensity weighting, selection bias transform, and the probability that each case would occur, given the off-road glance and maximum deceleration probability). These steps are described in each respective method description.

### 2.8.2 Validation of data selection bias

In addition to comparing the bias in the data through distributions and means of delta-v (the comparison across GIDAS, and the CBM and BLOM models), we performed two selection bias analyses. The first uses percentile plots: the delta-v of the original crash is compared to the distribution of generated crashes for that seed crash as a percentile. The second is an analysis of injury risk bias. We applied the injury risk functions for MAIS1+, MAIS2+, and MAIS3+ to the distribution of crashes from the CBM and to the same distribution transformed to mimic the GIDAS sample selection. The purpose is to compare the results, in order to investigate how the selection bias affects the calculation of injury risk.

The injury risk functions used were those for frontal crashes reported in Wang (2022; Table 1.1). Note that, to keep this part of the work less complex, we did not consider the combination of rear-end injury risk and frontal injury risk that are part of a rear-end crash. Instead, we only considered the frontal impact injury risk.

### 2.8.3 Assessing the safety of ideal, hypothetical driver monitoring systems

We demonstrate how to perform an assessment of DMSs with interventions. The mean delta-v, the delta-v distribution, and the crash avoidance rate were used to compare the simulation outcomes of the original SHRP2 off-road glance distribution with the simulation outcomes of the SHRP2cut3s, and SHRP2cut2s distributions. The crash avoidance rate was calculated as described below:

- Step 1: Case-by-case crash probability calculations:
    a) For each original seed, calculate the crash probability for the original SHRP2 off-road glance distribution (baseline) as the sum of the combined probabilities of all permutations of off-road glance durations and maximum decelerations leading to a crash.
    b) Calculate the crash probability of the modified 'cut' data set as the sum of the combined probabilities of all permutations of off-road glance durations and maximum decelerations leading to a crash.
    c) Calculate the ratio between the crash probability for the 'cut' data set and the baseline crash probability.
- Step 2: Crash avoidance rate calculation:
    d) Subtract the obtained crash probability ratio from 1 to obtain the crash avoidance rate for each seed.
    e) Sum the crash avoidance rates over all seeds and divide by the total number of seeds to get the total crash avoidance rate.

## 3 Results

### 3.1 Comparing delta-vs for crashes generated from GIDAS, CBM, and BLOM

The main validation in this study is based on a comparison of the estimated delta-v of the 103 original GIDAS crashes and the CBM-generated crashes. The comparison was made possible by transforming the latter, using Eq. 2, to mimic the GIDAS sample selection process (see Figure 6). In addition to the main validation (GIDAS vs. CBM) we also studied how the BLOM model results compared with GIDAS and CBM.

The top panel in Figure 7 shows that the mean delta-v of the original CBM crashes, without sample selection transformation, is 13.63km/h—much smaller than that of the GIDAS crashes (M=17.84km/h), with a very large proportion of crashes at low delta-v. This was expected, as the simulations should produce crashes across all levels of severity, without the bias present in in-depth crash databases.

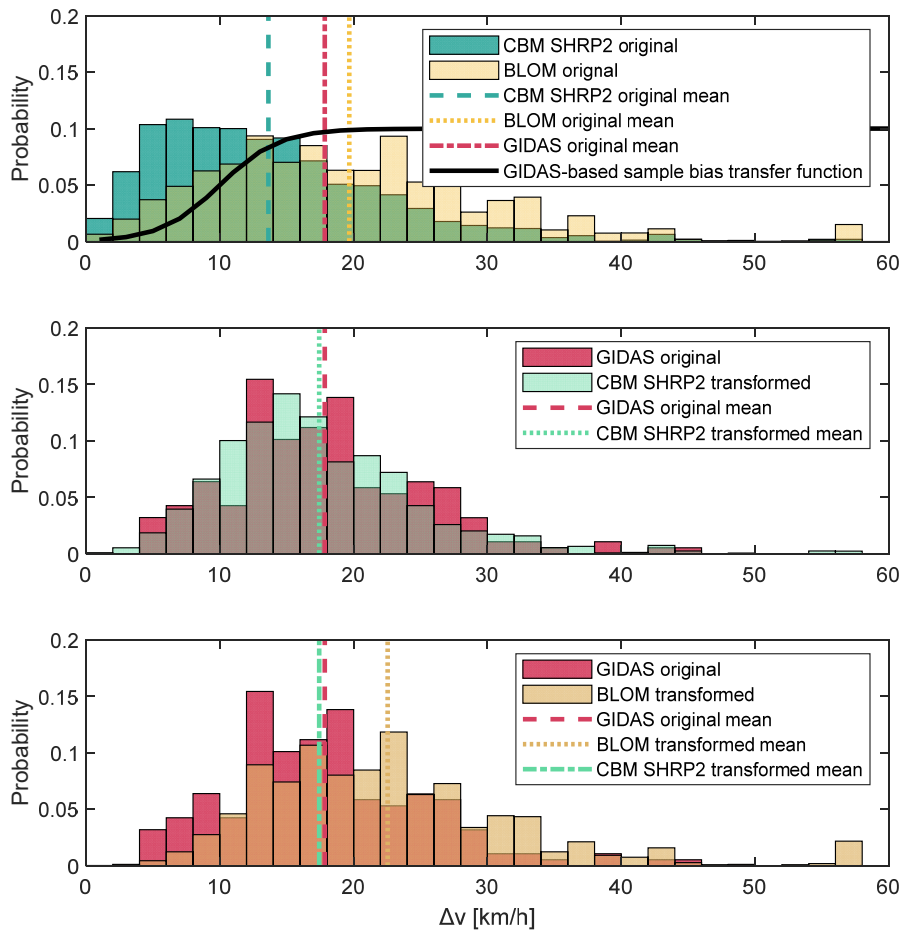

*Figure 7: A comparison between the original GIDAS cases and the scenarios generated using the proposed CBM and BLOM. The top panel shows the actual crashes generated by the CBM and BLOM (without transformation). The two lower panels show the normalized distribution from the top panel after Equation 2 (shown as the solid sigmoid in the top panel) is applied – that is, the distributions in the top panel have been transformed to mimic the GIDAS sample selection process (middle panel: CBM; bottom panel: BLOM).*

More important for model validation, however, the delta-v of the transformed distribution (M= 17.50km/h; middle panel, Figure 7) is now much more similar to that of the GIDAS distribution (M=17.84km/h), although still underestimating delta-v by 0.34km/h. We also compared the distributions using a range of summary metrics, available for enthused readers (see Appendix H).

In addition to comparing the CBM to GIDAS, we also compared it to a brake-light onset (BLOM) crash-causation model. The bottom panel of Figure 7 shows that BLOM substantially overestimates delta-v (M= 22.55km/h; a difference of 4.71km/h compared to GIDAS). Note that BLOM overestimates the mean delta-v even without the transform, and the non-transformed distribution does not include low-severity crashes, as it should. Also note that as the BLOM does not handle cases when the lead vehicle is not braking or is at a standstill, such cases (35 out of 103; 34.0%) were excluded from the BLOM simulations (and are thus not part of the bottom panel in Figure 7). See Appendix for more details on the comparison between CBM and BLOM.

**3.2 Percentile-based crash-causation model bias assessment**

The analysis in Section 3.1 is one way to investigate a model's validity in crash generation. However, it only demonstrates how well the overall delta-v distribution fits the validation data (GIDAS) and does not consider how well the crashes generated for each of the individual seed crashes actually represent the seed crashes. One way to evaluate how representative they are is to study the distribution of the percentiles of the delta-vs of the individual seed crashes in the distribution of the crashes generated for each respective seed crash. The method presented here was inspired by the work by J. Miller (2021) and discussions with Carol Flannagan (personal communication, November, 2023). Figure 8 shows an example of the percentile in the generated crashes for a specific seed crash; Figure 9 shows the histogram of the percentiles for all 103 crashes for CBM and the 68 crashes for BLOM (the subset of the 103 crashes for which BLOM is valid).

This type of validation is based on the fact that each individual seed crash is only one instantiation of the probability of a crash occurring (given the combination of all factors, here including the off-road glance duration and the deceleration). Thus each seed crash may be anywhere in the distribution of possible crashes for that seed, but the percentile histogram of the distribution (across all 103 crashes for CBM and 68 crashes for BLOM) should be uniform. That is, an individual seed crash can occur at any delta-v that is possible for the specific concrete scenario, but the distribution of the percentiles of all possible crashes should be uniform. This can be seen as the reverse of random sample generation. (See Appendix K for details.) Note that ensuring distribution uniformity only validates the scenario generation from the seed crash distribution. It is a separate issue, beyond the scope of this paper, to ensure that the sample of seed crashes chosen represents the real world.

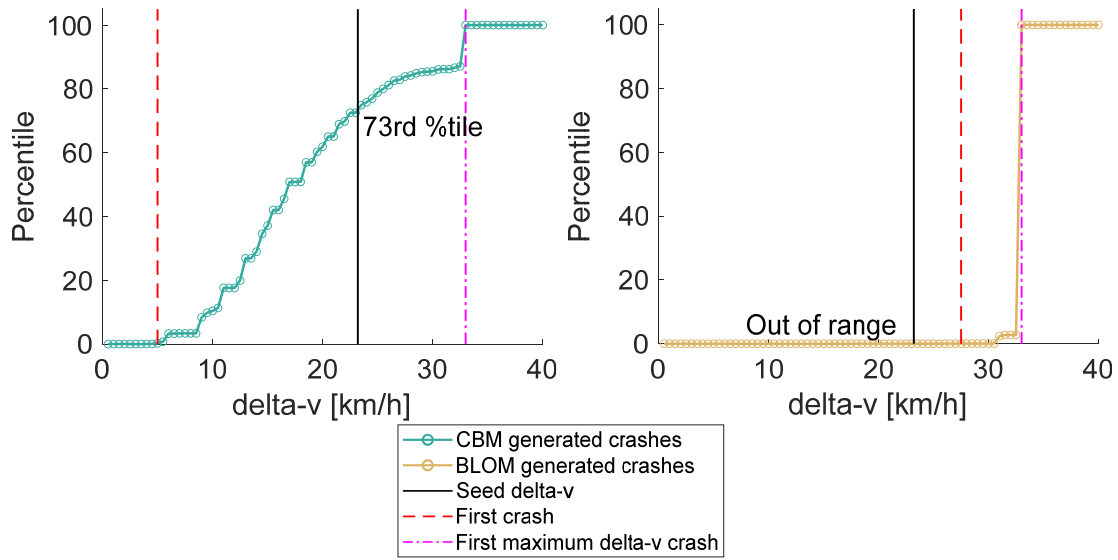

*Figure 8: One CBM and one BLOM example of percentile extraction. The left panel in Figure 8 shows that the seed delta-v was approximately 23km/h, and that it is percentile 73 of the delta-vs of the CBM generated crashes for this particular seed. The 'jump' at the top is the 10% added sleepy no-reaction drivers. The right panel shows the same seed, but where BLOM is applied, where the BLOM model does not brake until very late (i.e., the lead-vehicle brake lights are not turned on until shortly before the crash). This results in the seed delta-v (23km/h) being outside of the range of the BLOM-generated crashes for that seed (27.5-33km/h). The vast majority of the BLOM-generated crashes for that seed crash with maximum delta-v. The data at percentile 100 extend beyond 40km/h, but they are cut at 40km/h for consistency between panels and clarity.*

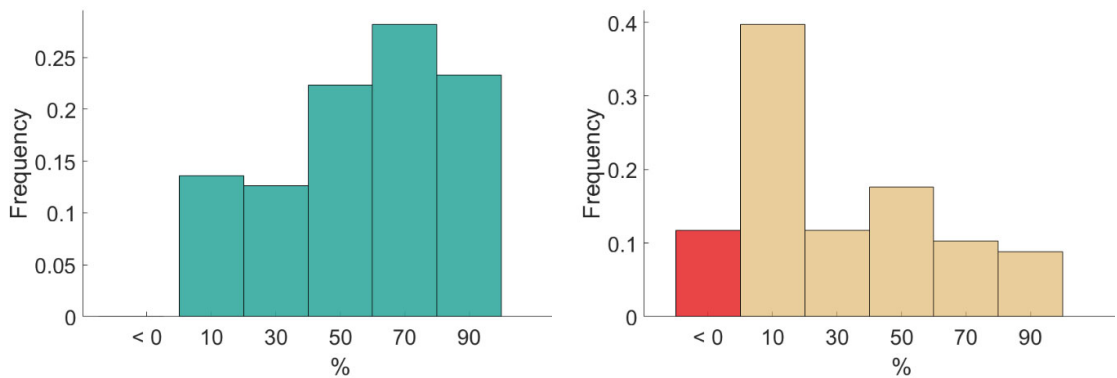

*Figure 9: A histogram of the percentiles of the individual seed delta-vs in their respective set of generated crashes for CBM (left) and BLOM (right). The red bin is when the delta-v of the GIDAS seed ended up outside of the delta-v distribution of the generated crashes for a specific seed crash. The left percentile example in Figure 8 is a datapoint in the 70-percentile bin in the left panel in this figure, while the right example in Figure 8 is one of the red datapoints in the right plot.*

If the crash-causation model is perfect, the percentile histogram should be completely uniform. If it is skewed to the left, the delta-vs of the simulated crashes are, in general, higher than those of the seed crashes (and vice versa if it is skewed to the right). As can be seen in both panels of Figure 9, the percentile histograms are not uniform for either CBM or BLOM. Some non-uniformity is expected, as the histograms are only based on 103 and 68 cases for CBM and BLOM, respectively (see Appendix K for a sensitivity analysis of bin sizes). However, CBM is more uniform than BLOM, even if it is clear from Figure 9 that there are more seed delta-vs at the 'right' part of the distribution of the generated CBM crashes (per seed). This difference between GIDAS and the CBM model is not obvious in Figure 7. The fact that it is not obvious in the delta-v distribution illustrates the importance of performing validation from several perspectives. For BLOM, Figure 9 shows that—for as many as 40% of the GIDAS seeds—the seeds are at the lowest 20-percentiles of the crashes generated with BLOM (actually, the sensitivity analysis in Appendix K (Figure K4) shows it is even worse than that). This, different from CBM, is in line with the substantial overestimate of delta-vs in Figure 7. Furthermore, BLOM has values below the range of 0–100. This means that the seed crash delta-v was lower or higher than the minimum and maximum for that proportion of seed crashes.

Further research is needed to quantify the link between non-uniformity and scenario generation validity, possibly adding some threshold metric for acceptable 'validity'.

### 3.3 Validation-data selection bias

We have proposed a process for considering the selection bias of the dataset against which scenarios generated through virtual simulations are compared. Here we further investigate the impact of ignoring the bias when estimating injury risks. Specifically, we investigate the differences in terms of injury risk for MAIS1+, MAIS2+, and MAIS3+, by applying the three injury risk functions to the non-transformed virtual simulations and to the transformed simulations and comparing the results. Based on the dose-response model (Kullgren, 2008), we calculated the estimated proportion of injuries at each injury risk level. That is, we determined the proportion injured at each injury risk level for GIDAS, for the original (non-transformed) CBM-generated crashes, and for the crashes generated by the transformed CBM. The calculation conducted is

$$I_{MAISX+} = \int_0^\infty R_{MAISX+}(\Delta v) \cdot h(\Delta v) \cdot d\Delta v$$

where $I_{MAISX+}$ is the proportion injured for the level of injury, $R_{MAISX+}(\Delta v)$ the discretized injury risk function for each respective level, and $h(\Delta v)$ is the empirical delta-v distribution (i.e., GIDAS, the original CBM, and the transformed CBM crashes). Figure 10 shows the components in $I_{MAISX+}$, while Table 2 shows the proportion of injured individuals for all combinations. The transformed CBM is more similar to GIDAS than the original CBM is, across all levels of injury risk. The original (without transformation) CBM underestimates the proportions by 13% (MAIS1+), 21% (MAIS2+), and 17% (MAIS3+) compared to GIDAS. The transformed CBM is slightly off on the proportion of injured individuals, with an underestimate of MAIS1+ (by 1%) and an overestimate of MAIS3+ (by 6%). However, its estimate of MAIS2+ is off by less than 0.1% compared to the GIDAS MASI2+ injury risk estimate. Note that BLOM with transformation overestimates the injury risks substantially (by 16% for MAIS1+, 29% for MAIS2+, and 119% for MAIS3+) compared to GIDAS – much more so than CBM.

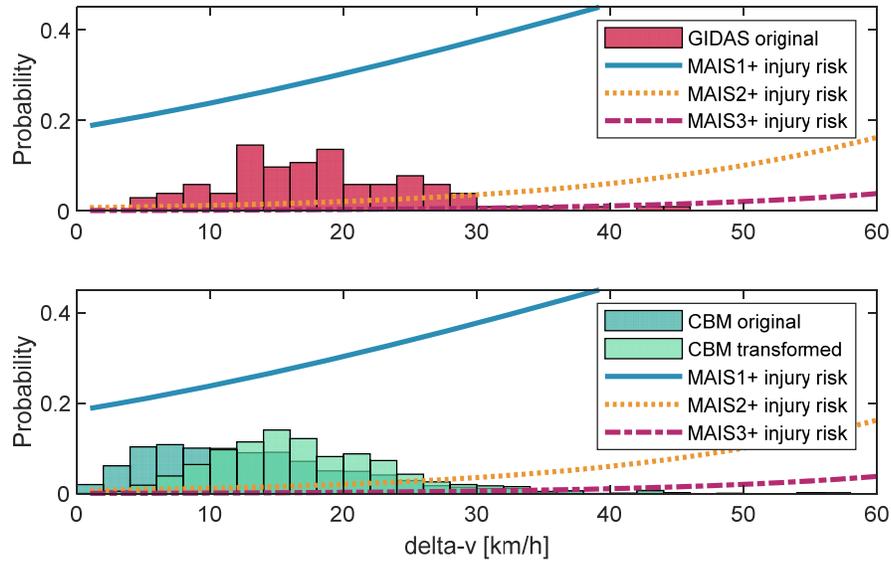

*Figure 10: A visualization of the injury risk functions ($R_{MAISX+}(\Delta v)$) and the models ($h(\Delta v)$).*

|  | $I_{MAIS1+}$ | $I_{MAIS2+}$ | $I_{MAIS3+}$ |
|---|---|---|---|
| GIDAS | 0.371 | 0.0412 | 0.00773 |
| Original CBM baseline | 0.323 | 0.0326 | 0.00642 |
| Transformed CBM baseline | 0.365 | 0.0412 | 0.00820 |
| Transformed BLOM baseline | 0.428 | 0.0698 | 0.0170 |

*Table 2: The proportion of occupants sustaining different levels of injury (when MAIS1+, MAIS2+ and MAIS3+ injury risk functions are applied) for three delta-v distributions: GIDAS, original CBM, and transformed CBM.*

### 3.4 Sensitivity analysis

We performed five sensitivity analyses, comparing the delta-v distributions between a) two different off-road glance duration distributions (see Figure 1), b) two different $\tau^{-1}$ thresholds (for glance anchoring; $\tau^{-1}= 0.2^{-1}$ and $0.3s^{-1}$), c) two different driver response times (0.5 and 0.8s), d) a range of different assumptions about the proportions of PDO crashes in "real traffic" (55% to 85% in 5% steps), and e) 18 variations of how the PDOs were added to the lowest bins (see Appendix G) and an example of what the results would be if the PDOs were log-normal-like. The results of the five analyses are presented in turn.

As the top panel of Figure 11 shows, the two glance distributions have a relatively large difference in delta-v distributions: the Kungälv glances shift the distribution towards lower-severity crashes by 1.4km/h. Studying the differences in detail (e.g., plots of delta-v, weights per seed crash) it seems that the reason for the large shift to the left for the Kungälv data is that the Kungälv off-road distribution does not include glances higher than 2.6s. This shows, as expected, that long glances, while rare, have an impact on the overall severity outcome.

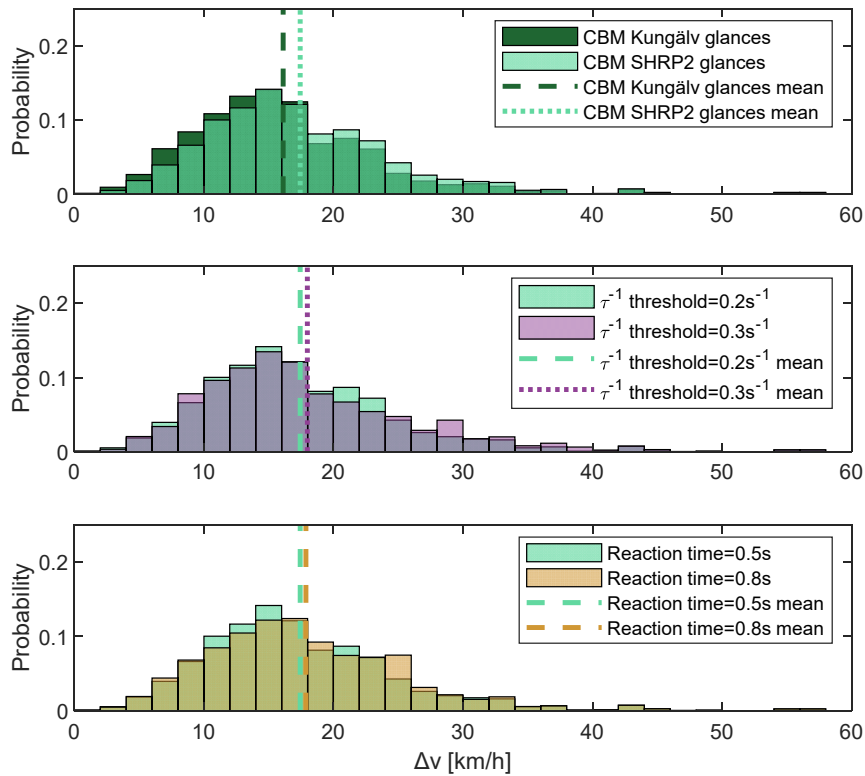

*Figure 11: Comparison of delta-v for the baseline scenario generation using the SHRP2, and the Kungälv baseline off-road glance distributions (top panel), the CBM with two different $\tau^{-1}$ thresholds ($0.2s^{-1}$ and $0.3s^{-1}$; middle panel), and for the CBM with two different response times (0.5s and 0.8s; bottom panel).*

As the middle panel of Figure 11 shows, the $\tau^{-1}$ threshold changed the delta-v distribution much less (mean change of 0.49km/h) than it changed the glance distribution.

As the bottom panel of Figure 11 shows, an increase in the driver response time from 0.5s to 0.8s shifted the delta-v distribution even less (mean change of 0.38km/h).

The fourth sensitivity analysis was to assess the impact of the choice of overall proportion of PDOs in the creation of the selection bias transfer function. The range 55% to 85% (i.e., +-15%) with a step size of 5% was assessed. The left panel of Figure 12 shows the results, which indicate that the mean estimate of the delta-v distribution after the original CBM is transformed is slightly sensitive to the choice of the total proportion of PDOs (used when selection bias transform function), with a range of approximately 2.6km/h difference in the mean when changing from 55% PDOs to 85% PDOs. In this study 70% PDOs were used when creating the selection bias transfer function for all other analyses.

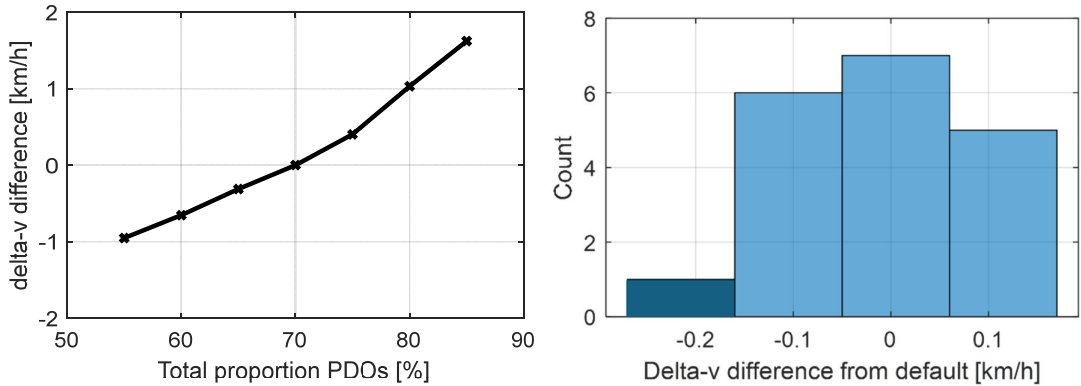

*Figure 12: Results of two sensitivity analyses. The differences between mean delta-vs of the transformed CBM for percentages of included PDOs ranging from 55% to 85% (left panel), and a histogram of the differences from the transformed CBM with 70% PDPs (best-fit; used in Figure 7 and throughout the paper) and the 18 variants.*

The aim of the fifth and final sensitivity analysis was to understand how sensitive the exponential fit, the transform fit, and the subsequent (using the transform) mean delta-v were to the way the PDOs were added manually to the lowest bins of the Folksam PDO (Figure 7) and to compare these findings to those of the GIDAS data (Figure 12). The right panel in Figure 12 shows a histogram of the mean delta-v difference compared to the "best manual fit" (used across this study), for a possible log-normal-like PDO distribution (leftmost bin) and 18 variants of the best manual fit (for each variant, each bin was randomly modified by +-30% of the best manual fit value). As can be seen, the difference for the random variants is less than +-0.2km/h, much less than the differences induced by changing the percentage of PDOs—and lower than the differences seen when assessing safety systems or glance behaviors. The log-normal-like distribution is a data point in the leftmost bin (dark blue), indicating that the transform is relatively insensitive to whether the distribution is log-normal or exponential. Further details of this sensitivity analysis can be found in Appendix G.

### 3.5 Assessing the safety of ideal, hypothetical driver monitoring systems

In this demonstration we use two different ways of comparing glance behaviors (behaviors that may be the result of an HMI, a nomadic device or the introduction of a DMS – here we demonstrate the method by assessing two ideal, hypothetical DMSs). The first way is to compare the means and shapes of delta-v distributions, as we have done previously in this paper. The second is to compare the crash avoidance rate—a metric of the DMS' crash avoidance performance. Note that the delta-v distribution is of the crashes that remains with the DMS (it excludes the avoided crashes). Also note that here the original (non-transformed) distributions are used, as we want to estimate the real-world delta-v (the transformation is only for the validation against GIDAS).

The top panel of Figure 13 shows the delta-v distributions of SHPR2cut3s compared to those of SHRP2. The delta-v of SHRP2cut3s (M=10.53km/h) is lower than that of non-transformed CBM SHRP2 (M=13.63km/h): a difference of 3.10km/h. The crash avoidance rate is 37%. Note that for the original SHRP2 glance distribution, all 103 seed crashes generated simulated crashes. However, for SHRP2cut3s, only 93 of the seed crashes resulted in simulated crashes;

almost 10% of the original crashes would not have occurred if drivers' off-road glances had never been longer than 3s.

The second panel from the top of Figure 13 shows the delta-v distributions of SHPR2cut2s compared to SHRP2. The delta-v of SHRP2cut2s (M=10.12km/h) is lower than that of SHRP2: a difference of 3.50km/h. The crash avoidance rate is 51%. For SHRP2cut2s, 30 seed crashes did not result in simulated crashes. Thus, if drivers had not made any off-road glances longer than 2s, almost 30% of the seed crashes would not have occurred—and, in total, approximately half of the crashes generated with the original SHRP2 glance behavior were avoided with the SHRP2cut2s ideal, hypothetical DMS.

We also estimated the injury risk reduction for the DMSs for those crashes that remained, using the same injury risk functions as before. Compared to the original SHRP2 CBM (all crashes), the SHRP2cut2s DMS reduced the MAIS3+ mean injury risk of the remaining crashes by almost 50%, that of MAIS2+ by almost 40%, and that of MAIS1+ by just above 10%. SHRP2cut3s actually reduced them almost as much (only a few percent less).

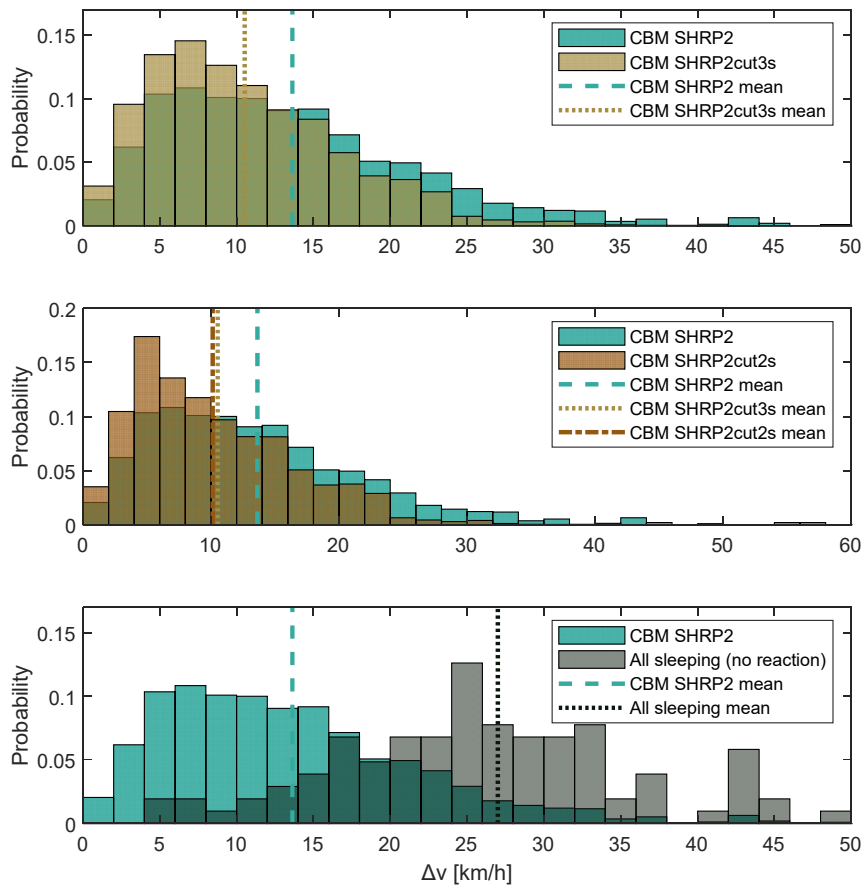

*Figure 13: Comparison of the delta-v between SHRP2 baseline glances and the two hypothetical DMS distributions (two upper panels), and the original CBM and all-sleeping-drivers (bottom panel).*

The bottom panel in Figure 13 is added as a reference. It shows what the distribution of delta-v would be if all drivers were sleeping (or not reacting for some other reason). That is, this distribution is the maximum (most right-shifted) delta-v distribution that the seed crashes can generate, given that the drivers would never accelerate into the crash. Here the mean is, as expected, much higher (M=27.00km/h) than any other. When the sleepy-driver glances (as described in 2.2.4) are added, this is the distribution that is added so that it makes up 10% of the overall crash distribution.

## 4 Discussion

In this work, the primary aim was to validate a crash-causation-research-based baseline scenario-generation model (called CBM), for use in pre-crash virtual safety assessment, by comparing the model's generated crashes against reconstructed crashes from the GIDAS database. The CBM has four sub-models, considering four different crash-causation mechanisms: Off-road glances (delayed braking response due to off-road glances), Too-close (the following-vehicle is driving too close to the lead vehicle), Low-deceleration (the driver of the following vehicle is not braking as hard as is possible under the circumstances), and No-response (e.g., sleepy or drowsy drivers). The validation was performed on the distribution of the estimated delta-v, as well as on summary statistics for the delta-v distribution and crash avoidance rates.

Because GIDAS has a substantial selection bias towards higher-severity crashes (and PDO crashes are excluded) while the CBM generates crashes at all levels of severity, it was necessary to devise a method to make valid comparisons between them. We created a transfer function that can transform a delta-v distribution of CBM crashes to a distribution with a selection bias similar to that of the GIDAS data. As noted, this step is important, as models like CBM typically produce crashes across all severity levels, while databases of actual crashes often have some form of selection bias. We also performed five sensitivity analyses to assess the model's robustness.

In addition, to demonstrate how driver monitoring systems (DMSs) with interventions can be assessed for their impact on safety the delta-vs for the CBM and two ideal, hypothetical DMSs were compared, and crash avoidance rates and the reduction in mean delta-v and mean injury risk for the two were calculated.

### 4.1 Comparing baseline delta-vs for GIDAS, CBM, and BLOM

Our results show that the proposed crash-causation-based scenario-generation model (the CBM with the selection-bias transform) creates crashes with a delta-v distribution similar in shape to that of the validation dataset (GIDAS), although the former is shifted somewhat toward lower-severity crashes. The estimated MAIS1+ and MAIS2+ injury risks for the transformed CBM are almost identical to those of GIDAS, with a slight overestimation of the MASI3+ injury risk (6%). Here the means were compared, but their similarity is relatively difficult to measure here, not least since most statistical tests are about quantifying differences, not similarity. Further, simulations generate many thousands of crashes—so any comparison, however similar the distributions are, will likely reveal significant differences. We do provide a variety of statistical metrics in Appendix H, but due to the issues described above we chose not to provide probabilistic statistics such as p-values.

That the CBM creates a relatively similar delta-v distribution and mean injury risk with sub-models based on crash-causation theory is encouraging, as a theory-based model without tuning facilitates transparency in all steps of the virtual simulation process. When the CBM is used to assess glance behavior (e.g., the impact of a DMS with intervention) or specific ADAS/ADS safety systems (e.g., AEB), it should be possible to analyze the benefits and drawbacks of the assessed behavior or system. In contrast, machine-learning-based methods, such as those described in Ding et al. (2023), do not offer at all the same level of transparency in the analysis. In-depth analysis, when possible, may reveal systemic issues and lead to improved solutions. One such example is the assessment of systems which include the driver as part of the system, such as forward collision warning systems (FCW; Benmimoun, Pütz, Zlocki, & Eckstein, 2013). Studies have shown that drivers do not initiate braking directly based on the warning (other than in artificial laboratory settings, when told to do so); instead the FCW actually redirects the driver's gaze to the forward roadway (Aust et al., 2013). When on- and off-road glances are explicitly part of the crash-causation model (e.g., the CBM), glance redirections can be explicitly included in the assessment; thus the warning can be modeled to interrupt off-road glances (as well as sleepy drivers). When there are no explicit crash-causation mechanisms (e.g., in machine learning-based baseline generation, as well as in purely stochastic baseline generation without simulations ; Wimmer, 2023) or reaction time-based models (such as BLOM), it is far from obvious how to assess systems (such as FCW or DMS) in which driver glance behavior plays a role.

We also assessed the performance of the BLOM, the simplistic crash-causation model which uses a simple reaction time (from a distribution) to the onset of braking of the lead vehicle. Results show that, compared to GIDAS, the BLOM model substantially overestimates the delta-v. Even more problematic is the fact that when calculating the injury risk for high-severity crashes, BLOM overestimates the MAIS3+ injury risk by more than a factor of two (~120%). This large disparity is not so surprising, however, as studies have shown that drivers' braking responses to critical events are primarily driven by urgency, and much less by the onset of a lead-vehicle's brake-light. Although brake-light onset explains some of the variance in responses, it is not a good predictor for the start of braking of the following vehicle in rear-end crashes (Markkula et al., 2016). Further, as previously mentioned, using brake-light onset is problematic (i.e., impossible) when the lead vehicle is neither braking nor moving. For these reasons the model could not be applied in 35% of the original crashes.

Appendix J describes a detailed analysis of the generated crashes per seed (i.e., the delta-v distributions), for all CBM and BLOM seeds. We found that the kinematics for CBM and BLOM are similar for 'easy-to-avoid' situations, when a low deceleration with a long glance (for CBM) or with a long reaction time (for BLOM) is needed to generate a crash. This means that BLOM may still have its uses, particularly when assessing safety systems specifically designed to avoid or mitigate the rare 'easy-to-avoid' situations. However, it should never be used when the lead vehicle never brakes at all or is at a standstill at the start of the event, because it is not even defined for these situations. Moreover, BLOM is typically not suitable for 'hard-to-avoid' car-following seed crashes either, because it will create substantial overestimates of the outcomes. Versions of BLOM that use a TTC threshold (for example) as the trigger, instead of brake light onset, may be more useful, but overall, CBM should be more universal and grounded in crash causation research.

### 4.2 Sensitivity analysis

We performed five sensitivity analyses in this study. The first assessed the impact of different baseline off-road glance duration distributions, showing that the choice of off-road glance behavior data does matter substantially. That is, as expected, when the SHRP2 off-road glance

data—with the longest glances being substantially longer than that of the Kungälv data—were used, the delta-v distribution was shifted towards higher values. Actually, we initially used the Kungälv glances as the main baseline, but when we realized that the length of the tail of the baseline matters, we switched to the SHRP2 baseline. This decision led to a better fit between CBM and GIDAS with respect to the delta-v distributions. One reason for a 'short' longest-glance may be the data size – if you have fewer data points (a low number of samples describing your distribution), you may not be capturing the rarest behaviors. It may also be that the drivers in the Kungälv study knew they were being observed (it was a controlled study on the road and therefore they may have been more focused on the driving task than they would have been under more naturalistic conditions). On the other hand, the SHRP2 data was fully naturalistic. To reiterate, care should be taken when choosing what glance baseline to use. Care should therefore be taken when using counterfactual simulations (with, e.g., CBM) with small sample sizes (empirical distributions). In fact, further research is needed to establish how sensitive it is to sample size. The design of the experiment must also be considered, as it may impact the way that drivers act during the experiment. If glance distributions can be modelled as Morando, Gershon, Mehler, and Reimer (2021) when they fit log-normal distributions to off-road glance behaviors of Tesla drivers, there is no issue with a "too-short" distribution (accurate modelling of the distribution is, however, required). For more complex distributions, such as when a DMS only affects longer glances, some form of mixed-model solution would be needed.

For the sensitivity analysis looking at the two model parameters and reaction time, the differences were smaller than expected. That is, when the CBM's $\tau^{-1}$ was increased from $0.2s^{-1}$ to $0.3s^{-1}$ and the reaction time was changed from 0.5s to 0.8s, the delta-v distributions were shifted only marginally (0.49 and 0.38km/h delta-v changes, respectively). That the changes were so small despite the increase in the total number of crashes (from 28,697 to 36,105) with the $\tau^{-1}$ increase and from 28,697 to 30,360 with the reaction time increase, is encouraging, as it indicates model robustness. Overall, these results indicate that the model is reasonably robust to changes in the a) glance distribution (although care should be taken with "too short" tails of empirical glance distributions), b) glance anchor ($\tau^{-1}$) threshold, and c) driver response time.

**4.3 The importance of considering sample selection bias in validations**

Our results clearly demonstrate how the sample selection bias of the dataset against which simulation-generated baselines are validated can affect the validation's accuracy and precision (see Hautzinger et al., 2005 for examples of GIDAS selection biases). Thus, performing validation using delta-v (or the equivalent, such as relative speed at impact) is clearly not advisable unless the selection bias is compensated for, as we propose. Otherwise, any simulation-generated baselines that show good correlation with, for example, GIDAS, would include many more higher-severity crashes than the true distribution.

Our results also show that using injury risk functions as a transform instead of a specifically designed selection bias transform (such as what we propose here), substantially underestimates the injury risks. As noted previously, our hypothesis was that using injury risk functions for higher injury risks (e.g., MAIS3+) as a 'selection bias transform' would be better (create a transformed distribution closer to GIDAS), but it turned out to be the opposite. That is, applying MAIS2+ and MAIS3+ functions as 'GIDAS selection transforms' (rather than an explicit selection bias transform – what we do in this paper) generated delta-v distributions with substantial underestimates of injury risks than when using MASI1+. We expected the "bend" in the injury risk curve to basically remove the low severity crashes (similar to the transform we developed), but looking at Figure 10, all three injury risk curves are still relatively flat where there is data in the delta-v distribution. This flatness of the injury risk curves over the range of the delta-v data in this dataset is likely the reason why the $I_{MAISX+}$ for all three seems to be underestimating the injury risk. This will differ much between scenarios. That is, rear-end crashes typically have low risks of injury. It may be that using an injury risk function as a proxy for an explicit selection bias transform in for crashes with higher risk (e.g., crashes with vulnerable road users) will reduce the underestimation. However, this should be assessed in further research.

The second type of validation we performed, studying percentile plots, has not been previously applied in the domain of virtual safety assessment. By studying where the seed crash delta-v is in the delta-v distribution of the simulated crashes, across all seed crashes, we can identify systematic biases in the crash generation process. Further exploiting this information to explore other aspects of model validity is beyond the scope of this research and deserves dedicated research efforts. It is worth mentioning that the validation method proposed here is only applicable to counterfactual simulations (baseline type B in Wimmer, 2023) in which a seed crash is used as a basis, and not in simulations that are completely model-based (without an anchor in individual original crashes; baseline type C in Wimmer, 2023).

As validation of crash-causation models for virtual safety assessment is rare, there is not much literature on the importance of considering sample selection bias in the validation of virtual safety assessment. Our results show that this approach mitigates the issue but does not completely solve it. There is research on selection bias using crash databases (e.g., Blincoe et al., 2023; Hautzinger et al., 2005), but not in the context of validating virtual simulations. Further, a recent study by Di Lillo et al. (2023) describe the importance of comparing the datasets with the same range of outcome severity in ADS assessment – supporting the need for proper transformations when datasets do not have similar selection processes with respect to outcome severity.

### 4.4 The potential of tuning the CBM

A relatively unique aspect of the crash generation process proposed through the CBM is that the model was not tuned to fit some outcome—instead, crash-causation theory and relevant knowledge and data about driver behavior were used to create the model. This information was enough to provide reasonably good validation results. It would certainly be possible to tune the model by tweaking its parameters (such as the $\tau^{-1}$ threshold or the reaction time), but then it would no longer have a theoretically accurate basis. However, as the proportion of PDOs in real traffic (used in the creation of the sample selection transfer function) and of no-reaction crashes have large uncertainties, it may be that the corresponding parameters should be assigned values other than those used in this study. The model could potentially be tuned on these variables, modifying their values to reach a specific outcome—as long as realistic values are used. For example, if the percentage of no-reaction drivers is increased from 10% to 13.5%, the mean delta-v is identical to that of GIDAS. A shift towards larger percentages of PDOs would also shift the distribution closer to that of GIDAS. A combination of the two may, however, be a better approach. Even if this is possible, further investigation into how to perfect the GIDAS selection bias transform should be done first—paying particular attention to the underreporting of low-severity crashes.

**4.6 Assessing the safety of a hypothetical driver monitoring system (DMS)**

This work did not aim at deep diving into the details of distraction and its implication on safety but set out to validate a crash-causation-based scenario generation model and demonstrate how virtual safety assessment can use the model to assess driver glance behavior. There are some previous work assessing glance behaviors through virtual simulations (Bärgman et al., 2017; Bärgman et al., 2015; J. Y. Lee et al., 2018), but previous studies have not been applied to crash-database data (instead lower-severity crashes from naturalistic driving studies were used). Further, the current study used a more complete crash-causation model than what can be found in literature. In particular, this is the first time that prevalence weighting is used. Taken together, the proposed method is substantially more complete and ready for practical use, compared to previously existing methods. This also means that the assessment of the DMSs is likely to be relatively accurate, and that future real DMS data could be "plugged in". To this end, proper baseline (glance behavior data from driving without a DMS) and treatment (glance behavior data from the same conditions as for the baseline, but with the DMS) would be necessary. If the DMS is added in combination with some crash or conflict avoidance system (e.g., AEB or ADS) , those should also be included in the assessment (see, e.g., Bärgman & Victor, 2020 for an example of such a combination).

The results of the analysis of the ideal, hypothetical DMSs (or, actually, the resulting off-road glance distributions of such systems) were as expected: Shortening glances reduces both the number of crashes (i.e., crash avoidance) and lowers the delta-v for the crashes that are left.

Although this study includes substantial methodological improvements for both scenario generation and its validation, compared to previous work, there are still several limitations and future research needed.

**4.7 Limitations and future research**

The limitations have been divided into five categories: the original GIDAS data representativeness, the creation of a selection bias transformation function, the CBM model, crash causation research, and the DMS assessment.

We assumed that the GIDAS data used are representative of the rear-end crashes that the crash-causation model is designed to generate. This assumption is important, because this work focuses on the challenges of validating crash generation in simulations. However, ensuring the validity of the samples of a specific scenario type in crash databases (or a subsample thereof) is a complex topic that has its own body of research (e.g., Hautzinger et al., 2005; Liers, 2018; Fan Zhang & Chen, 2013; F. Zhang, Noh, Subramanian, & Chen). In preparing this study we decided that this large topic is beyond its scope, since we have focused on validation aspects of the process after seed crashes are selected. Any difference between the seed crashes used and the 'true' (real-world) distribution would have an impact on the comparison of the CBM and BLOM models to the 'validation data' (the seed crashes), but such a difference cannot be quantified without a deep investigation into the seed crash representativity. Future work on the validation challenges in seed selection is needed to complement this work on the validation challenges 'post-seed-crash selection'.

There are another two main limitations related to the GIDAS data used in this study. One is that we used a limited number of GIDAS crashes for validation. Having more cases would make the validation assertions firmer (and getting more out of the DMS assessment comparison). Another limitation, inherent to the GIDAS data (and all reconstructed pre-crash kinematics), is that the PCM data are just reconstructions, with assumptions about the pre-crash behaviors (although they are the best possible assumptions, given the available data; Hautzinger et al., 2005; Otte, 2003). Deceleration profiles and speeds are therefore best estimates, rather than measurements. Specifically, crashes are reconstructed without much information about the initial speeds of the involved vehicles, the actual timing of braking, and the actual level and shape of the deceleration (i.e., jerk and maximum deceleration). Instead, these values, such as the initial speed and deceleration, are assumptions. Even if the GIDAS estimates of delta-v are reasonably good, the initial conditions in the simulations are based on consequential assumptions, not measured data. As a result, the crash-causation models presented here might be correct, and the uncertainties in the GIDAS pre-crash data make the difference (i.e., that the crash-causation models create crashes based on uncertain initial conditions and lead-vehicle braking behavior). Appendix I provides further details on the GIDAS data in relation to crash-causation factors. Future work should try to use more realistic deceleration (timing and shape) data, such as what is proposed in Wu, Flannagan, Sander, and Bärgman (In press).

The second limitation category—the creation of a transfer function that can be applied to crashes generated by, for example, the CBM—includes four primary limitations. The first is that we do not know the shape of the complete set of PDOs (across all severities; the Folksam data we used have an inclusion criterion based on repair costs). In this work we assumed that the distribution shape exponentially decreased with delta-v. However, it may instead be, for example, a gamma distribution. Optimally, future work would acquire data on the relationship between delta-v and repair costs in order to determine the distribution's correct shape. (In addition, the manual adding of data can be automated, although it is not likely to improve the model or model fitting substantially.) Second, the Folksam data used here include all kinds of frontal crashes. The PDO distribution is likely to be different if only rear-end crashes are included (given the GIDAS inclusion criteria used in this study). However, the distribution's shape was mainly used to create our transfer function (Eq. 2); we argue that it is not unreasonable to assume that the shape will be similar across many (if not all) scenarios. The third limitation is also related to the Folksam data: we only have data on injuries at the individual level, not at a per-crash level. This lack of information adds uncertainty to the estimates of the absolute values of the PDO distribution. Again, the fact that we primarily use the shape of the delta-v distribution should mitigate the impact of this limitation on the parameter fitting of the transfer function (Eq. 2). The fourth and final limitation is that the underreporting of low-severity crashes in GIDAS was only partially considered in the generation of the transfer function. Further research is needed to compensate for the underreporting in the best way possible (available data on underreporting has very low 'resolution', in that weights are typically only available for low- and high-severity crashes, making compensation based on them problematic).

As for the third limitation category—limitations of the CBM model—we acknowledge that it will not create all types of rear-end road crashes. That is, the causes of every crash are to some extent unique, and our model does not include more than a fraction of all possible causes of crashes. However, we argue that the CBM is likely to generate crashes representing the vast majority of rear-end crashes on highways. Although it is possible to develop crash-causation sub-models that represent even more crashes, the contribution to the overall pool of generated crashes from each added sub-model is likely to be small. Notably, though, the No-response sub-model should be further investigated. We used 10% No-response drivers due to sleepiness. Not only should the percentage used be anchored more firmly in literature, but there are also several other factors (in addition to off-road glances and sleepiness) that may result in No-response crashes. See Appendix F for more discussion on that topic. Further, the maximum deceleration of the following-vehicle driver (as seen in Figure 3) is from SHRP2, which primarily includes low-speed, low-severity crashes. However, measured data on the deceleration responses to crashes is rare: EDR data with this information often has a sampling frequency that is too low to accurately estimate maximum (plateauing) deceleration. Future studies should try to characterize a more representative deceleration distribution for the following vehicle, in the same way that lead-vehicle braking has been characterized by Wu et al. (In press).

Another limitation of CBM is that we do not consider context in the glance distribution used in the model. Although several studies have investigated details of the relationship between driver glance behavior and crash-causation (J. D. Lee et al., 2002; Markkula et al., 2016; Victor et al., 2015), few have investigated how context, such as the distance to the lead vehicle, impact the off-road glance distribution. A study by Bianchi Piccinini, Engström, Bärgman, and Wang (2017) indicates that drivers in a driving culture with short time gaps in traffic (e.g., China) keep their eyes on the road much more than drivers in a culture with (typically) longer time gaps (e.g., the US). This finding is likely also true within a driving culture: that is, when drivers drive with a short time headway, their eyes are likely to be on the road more than with longer time headways. Future studies should create contextualized glance distributions which should be incorporated into simulations that assess glance behavior.

One limitation of CBM relates to the No-reaction sub-model: in contrast to the other three, it cannot easily be integrated into simulations of continuous driving (which does not consist only of short-duration events) such as traffic simulations. Methodological work on how to integrate it into the main model is needed.

Further, there are certainly several areas where more crash-causation model research (the fourth limitation category) is clearly needed. Specifically, human behavior-related crash-causation models are needed for ADS (e.g., for take-over-requests) and ADAS, as trust in the systems increases and expectation mismatches become more common. Mismatch studies such as Bianchi Piccinini et al. (2020) are important. The Victor et al. (2018) study found that approximately 30% of the drivers participating in the study crashed into objects in front of them, even when they had their eyes on the road; the drivers expected the vehicle system to avoid the crash automatically. However, it is not obvious to computationally model trust and expectations. More research is needed to pursue such modeling.

Further, our crash-causation model only covers rear-end car-to-car crashes. Much more work is needed to create and validate crash-causation models for other scenarios, such as car-to-vulnerable road users (e.g., cyclists and pedestrians). The starting point for developing behavior-related crash-causation models for other scenarios should be a study of crash-causation for the scenario in question. Examples of such studies include those (Bianchi Piccinini et al., 2017) and Habibovic, Tivesten, Uchida, Bärgman, and Ljung Aust (2013), which used data recorders with video (dashcams) as a basis for the thorough analysis of crash-causation. Based on such studies, the crash-causation mechanisms need to be operationalized as computational models, which requires existing data and specific studies (such as driving simulator studies). However, the crash-causation model alone is not enough. Critical event response models (i.e., how the driver responds to a crucial event) are also needed, either pre-existing or developed specifically for the simulations. Jointly the models can then be applied to pre-crash kinematics data, such as GIDAS PCM. Further, new transforms to consider the missing PDOs and low-severity selection bias need to be developed for each new scenario (or, at least, the validity of the transform proposed here should be assessed before being applied to other scenario categories).

There is also one main limitation with the DMS assessment (the fifth and final limitation category): we are only demonstrating the concept of assessing a DMS. The DMSs are assumed to be ideal, similar to the way that ideal sensor and system performance is assumed when assessing an AEB system. Sander (2017) estimated that using ideal sensors when assessing an AEB may overestimate the benefits by 50%. However, as the scope of this work is to demonstrate how DMSs can be assessed, it should be acceptable to use ideal DMSs in this study. That said, it is not obvious what an ideal DMS would look like. If a DMS were to cut away *all* glances longer than a specific off-road glance duration, there might be safety-critical situations where longer glances are actually warranted from a safety perspective—in which case removing them would impact safety negatively. This possibility is for future studies to investigate.

## 5 Conclusions

This work set out to validate a theory-based, rear-end crash-causation model (CBM), consisting of four sub-models, for scenario generation. Although the shape of the delta-v distribution of the generated crashes turned out to be similar to that of the GIDAS data, the former underestimates the delta-v: the distribution is slightly shifted towards lower-severity crashes. However, the estimated mean injury risk for CBM was almost identical (within 1%) to the GIDAS data for MAIS1+ and MAIS2+, while it was slightly overestimated for MAIS3+. We also compared the CBM with a much simpler model (BLOM, the brake-light onset-based model), which substantially overestimated both delta-v and injury risk—the latter by more than 100% for MAIS3+. Also, the BLOM is not defined for 35% of the original crashes (when the lead vehicle does not brake or is standing still during the entire event). Further, recent literature does not support brake-light onset followed by a reaction time as a model for driver responses in rear-end crashes (Markkula et al., 2016). We argue that using CBM or similar crash-causation-theory based methods is more likely to capture both benefits and drawbacks of any technology or behavior change under assessment.

Results from our sensitivity analysis showed that CBM was relatively sensitive to the choice between two off-road glance distributions, but only marginally sensitive to the variation of two model parameters ($\tau^{-1}$ and reaction time), indicating that the CBM is a relatively robust crash-causation model with respect to the two latter parameters. As for the glance behavior, care must be taken when choosing which baseline behavior to use in CBM. While model robustness may be desirable, the validation will never be better than the underlying data, where selection bias is a genuine concern.

Although selection bias in crash databases is a common research topic, no work on selection bias related to the validation of virtual scenario generation appears to exist. The proposed method to compensate for the GIDAS (or other crash database) selection bias in the validation (by transforming the generated scenarios using models based on insurance data) seems to work reasonably well. Further, the selection bias transfer function seems to be relatively insensitive to the proportion of property-damage-only (PDO) crashes in real traffic and quite insensitive to the shape of the distribution at the lowest delta-vs. Note, however, that, a thorough validation of this method is difficult and will require more work. Specifically, more research on the design and validation of selection bias transforms for scenario generation is needed to: a) determine the shape of the full PDO distribution in the real world, b) extend it to and test it on different scenario categories (e.g., car-pedestrian interactions and car-car interactions at intersections), and c) validate the method itself (making sure that the validation method itself is accurate and robust).

The results show that delta-v distributions from different source data should not be compared without considering any possible selection bias in the data set(s). Clearly, the validation of scenario generation depends on being able to compare comparable data. Our work also shows that merely performing a validation on the injury risk (i.e., applying an injury risk function to both the validation data and the generated data) without addressing selection bias is not advisable, either. Doing so is likely to result in substantial underestimation of injury risk.

This work demonstrates that, in order to validate scenario generation for virtual safety assessment, there is a clear need to understand the details of scenario generation, as well as of the validation dataset used—especially with respect to possible selection bias in the validation dataset. Research on methods for validating scenario generation for virtual assessment is rare; much more work is needed for virtual simulation to become the tool we all want it to be, providing accurate estimates of safety benefits across all levels of crash severity. Finally, this work also indicates that behavior-based crash causation models may be an efficient tool to assess systems that aim to influence behaviors, such as DMSs.

**Sponsor statement**

The findings and conclusions of this paper are those of the author(s) and do not necessarily represent the views of the SHRP2, the Transportation Research Board, or the National Academies.

**Acknowledgements**

We want to thank the sponsors of this work, which include the QUADRIS project, funded by the FFI program; and the V4SAFETY project, funded by the European Commission under grant number 101075068. We also want to thank Sava Iancovici for supporting on data extraction, Xiaomi Yang for supporting in the simulation implementation. Also Ulrich Sander, Carol Flannagan, and Linda Boyle deserve our gratitude for valuable discussions on part of this work. We also want to thank Anders Kullgren at Folksam, who made the insurance data available for this analysis. The SHRP2 data used in this study has the identifier DOI SHRP2-DUL-16-172 and was made available to us by the Virginia Tech Transportation Institute (VTTI) under a Data License Agreement. Finally, we want to thank Christina Mayberry for her language review.

**Appendix A – The rationale for the assumption of 70% PDO for rear-end in real-traffic**

Knipling et al. (1992) stated that in the US in 1990 there were 1.5 million police-reported rear-end crashes on roadways with roughly 800,000 associated injuries and 2,078 associated fatalities. It is also approximated that there is an additional, unreported number of PDO rear-end crashes total between 1.3 million and 1.75 million annually (depending on methodology used). This gives a range of a total of 2.8 million to 3.25 million rear-end crashes, with 800,000 of these incurring injuries and the rest PDO crashes. Expressed as a percentage, the PDO crashes account for approximately 71% to 75% of all rear-end crashes.
More recent research (Blincoe et al., 2023) estimates that almost 74% of all crashes are PDO crashes (10,893,719 PDOs out of 14,193,727 crashes in the US in 2019; Tables 1–3 in the Blincoe report). The report also indicates that approximately 60% of PDO crashes and 30% of non-fatal injuries are not reported to the police. In this paper we use 70% to represent the percentage of PDO crashes.

**Appendix B – Estimation of delta-v**

The impact speed for the following vehicle from each simulation is estimated by assuming a one-dimensional inelastic collision between the following and lead vehicles. Due to the nature of car crashes in general (kinetic energy is not conserved), as well as the specific conflict situation studied (rear-end crashes), this estimation is shown to be reasonable, as seen in Appendix E.

Figure B1 shows a diagram of the lead and the following vehicles in an inelastic collision. The impact speed for the following vehicle from each simulation is then estimated as the change in speed for the following vehicle before and after the collision.

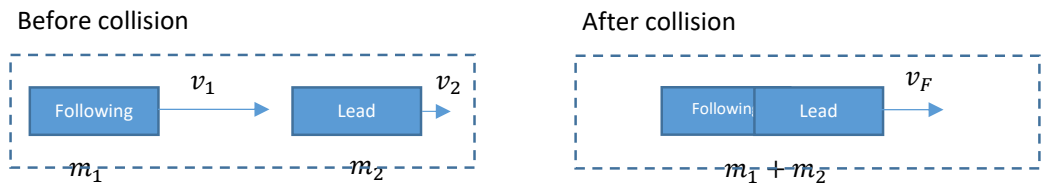

Figure B1: Diagram of involved vehicles in an inelastic collision

Conservation of momentum in the system before and after the collision yields Equation B1. The speed reduction for the following vehicle, $\Delta v_1$, is calculated according to Equation B2.

$$m_1 v_1 + m_2 v_2 = (m_1 + m_2) v_F \qquad (B1)$$

$$\Delta v_1 = v_1 - v_F \qquad (B2)$$

Solving Equation B1 for $v_F$ and substituting into Equation B2 yields:

$$\Delta v_1 = v_1 - \frac{m_1 v_1 + m_2 v_2}{m_1 + m_2} = \frac{m_2(v_1 - v_2)}{(m_1 + m_2)} \qquad (B3)$$

In practice, $v_1$ and $v_2$ are taken from the simulation output at the point in time when the following and lead vehicles first overlap, and $m_1$ and $m_2$ are taken from the corresponding crashes in GIDAS. Note, as seen in Equation B3, if the involved vehicles have the same mass, then $\Delta v_1$ can be estimated as simply $\frac{v_1 - v_2}{2}$. This estimation, also studied in Appendix D, can be useful when information is lacking about the relative masses of the involved vehicles.

**Appendix C – The overshot distribution**

The part of the glance that extends beyond the $\tau^{-1}$ threshold is called the overshot. If one assumes that the probability that any driver starts to look off-road is independent of the critical event initiation (e.g., that a lead vehicle brakes hard), it is possible to create what is called an overshot distribution, placed at $\tau^{-1}=0.2s^{-1}$. To exemplify, consider the driver's distribution of off-road glances shown in Figure 1 and imagine an off-road glance duration of 0.3s. This glance can be placed so that all of it overshoots the anchor point $\tau^{-1}=0.2s^{-1}$, or 0.2s of it overshoots, or 0.1s of it overshoots.

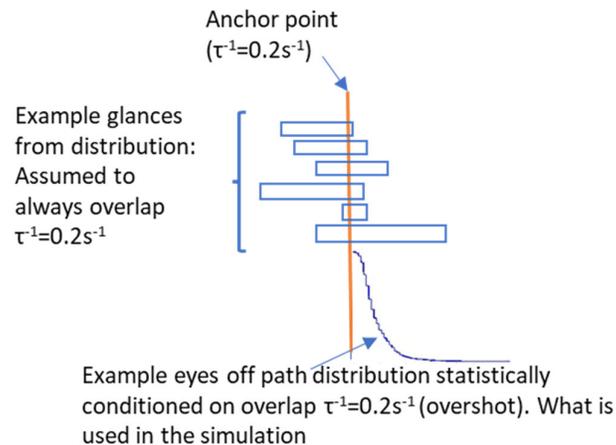

*Figure C1: Illustration of the overshot concept*

With the assumption that there is a uniform probability that the critical event will occur at any time during the 0.3s, those three glances have an equal probability of occurring (i.e., 1/3). For an off-road glance of 0.4s, overshoots of 0.1, 0.2, and 0.3s, and even 0.4s can occur. Now each of the four durations has a probability of 1/4 of occurring. For a 0.2s glance, only 0.1 and 0.2 can occur (with a probability of 1/2), and for a 0.1 glance only 0.1s can occur (with a probability of 1). This means that in terms of the overshoots of an entire off-road glance distribution, the probability of a short glance overshoot is much higher than the probability of a long glance overshoot (consider the sum of probabilities for 0.1s compared to 0.4s in the section above). There is actually a statistical transformation that can be performed (instead of running all these simulations) to convert a probability density function of off-road glance durations into an overshot distribution. See Extra Material for the code for the transformation. The benefit of this transformation is that it can be placed at the overshot anchor, which here is $\tau^{-1}=0.2s^{-1}$. See Figure C1 for an illustration of the overshot concept. This action reduces the dimensionality by one. That is, simulations are no longer needed for all combinations of possible overshoots. For example, instead of having to run 4+3+2+1=10 simulations for the 0.4, 0.3, 0.2., and 0.1s off-road glance durations, only four simulations are needed (one for each off-road glance duration). In our case, the transformation reduced the number of simulations needed by a factor of approximately 30 (that is, 30 times fewer simulations were required for SHRP2 baseline).

Further, the rationale for using the anchor at $\tau^{-1}=0.2s^{-1}$ comes from Markkula et al. (2016), Figure 2, where there is a really distinct threshold at $\tau^{-1}=0.2s^{-1}$. When drivers look back on the road after $\tau^{-1}=0.2s^{-1}$ they react directly (very quickly). Before $\tau^{-1}=0.2s^{-1}$ they may wait some to respond. This is an indication that drivers would not look off-road again if they have reached $\tau^{-1}=0.2s^{-1}$, which means that the last glance that is relevant for crashing is the one that is overlapping with $\tau^{-1}=0.2s^{-1}$. Hence the overshot anchor at $\tau^{-1}=0.2s^{-1}$

**Appendix D – Simulation Reduction Logic**

"Binary search" logic has been used to reduce the number of simulations needed in this study. The binary search algorithm is commonly used to perform an efficient search for an item in a sorted list. It starts in the middle of the search space, determines whether what we are searching for is above or below the current value, and then removes the half of the data which doesn't contain the sought item. It repeats the process until the item is found.

Assuming that there is a linear relationship between off-road glance length and impact speed, we can start by applying this logic and simulating the middle glance of a distribution. Based on the outcome (impact speed), we continue searching the distribution for a shorter or longer off-road glance length until we have found the shortest off-road glance that crashes. When it is found, we can simulate all subsequent off-road glances (skipping the ones for which we already know the results) until we have two subsequent crashes with the same impact speed, at which point we stop simulating. We then know which simulations we have not performed, and we can add them (non-crashes and crashes at maximum impact speed) without actually having to run them.

**Appendix E – Validation of delta-v**

*Estimation of change in velocity, $\Delta v$, during the collision*

The goodness of the estimation of $\Delta v$ by the chosen method was assessed by applying the method to the PCM data from GIDAS, as described in the method. For the PCM data we used the last data point, which corresponds to the index of collision, and applied the same equations. This process was performed for the 103 cases which fit the selection criteria. Figure E1 (left) shows the estimated impact speeds vs the impact speeds reported in GIDAS; the estimate was shown to be very good for most of the cases. Figure E1 (right) shows that if one does not consider the relative masses of the vehicles, the estimation is less accurate.

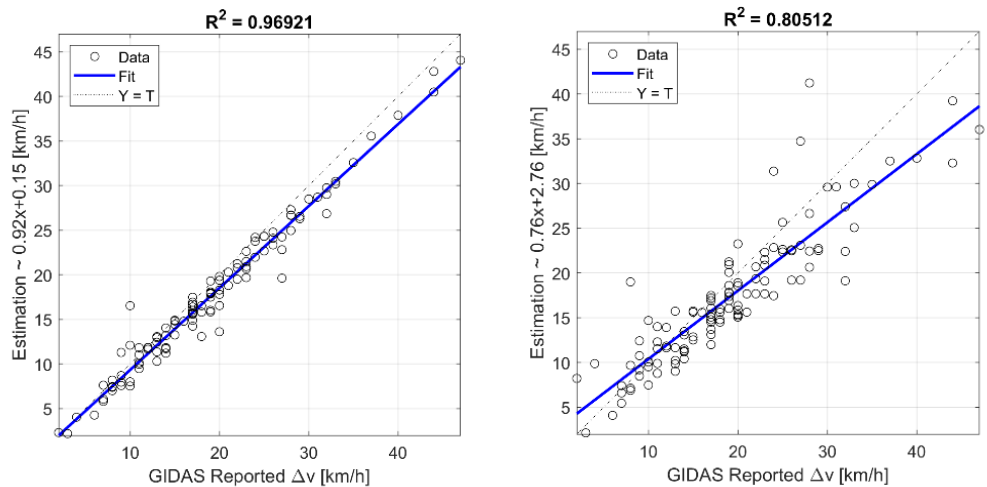

*Figure E1: Linear regressions of the relation between the $\Delta v$ reported in GIDAS and the estimation used in the paper (left) and a simpler estimation that does not take the relative vehicle masses into account (right).*

Since the chosen method still has some errors, the effect of these errors on the impact speed distribution was also analyzed—by comparing the impact distributions in the GIDAS original cases with the distribution of the estimated impact speed, using the selected method. Figure E2 show the magnitude of the effect of the impact speed estimation on the impact speed distribution.

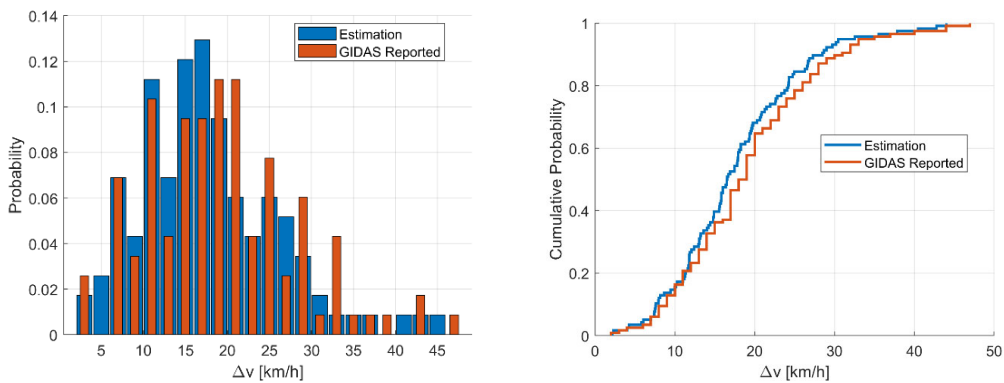

*Figure E2: Impact speed distribution difference between GIDAS original and the estimation (left), and the corresponding cumulative density function (right).*

**Appendix F – Simulated crashes per seed crash**

This section provides details about the generation of simulated crashes for each individual seed crash. It shows how the proportion of simulated crashes differs across seed crashes, both before and after the propensity weighting. It also illustrates what the proportion of maximum delta-v crashes is out of all crashes per seed. That is, it gives the proportion of simulated crashes per seed which crash with the maximum delta-v.

Note that what is shown here does not include the sleepy driver (no-reaction) crashes, as they are added at the very end of the process (per definition, since they are added on the histograms, after all other weighting is completed).

Figure F1 shows how the CBM crash generation works. The upper panel shows that although seed crashes with the fewest crashes are lower than the ones with the most crashes, they are "only" lower by a factor of two. That is, if the propensity weighting were based only on this, the maximum weights would be in the order of two (2). The lower panel shows the probabilities of crashing for each seed crash, which takes the weight in the upper panel into account while considering the probability that each generated crash will occur (given the probability of deceleration and off-road glance behaviour). The rightmost seed has a very small probability of resulting in a crash: the longest glance and the lowest deceleration would both have to occur for there to be a crash. The range between the highest and lowest probabilities in the lower panel is on the order of 21000.

As described in the main paper, the unreasonably large different in weights (the inverse of the probability) is handled by allowing a minimum and maximum weight of the 5$^{th}$ and 95$^{th}$ percentiles.

Note the very large difference between the upper and lower panel of Figure F1 in proportion of crashes that crash with maximum delta-v. The probabilities of crashing with maximum delta-v, when the deceleration and glance probabilities into account, is very low. All sees have some crashes that crash at maximum delta-v, also in the lower panel. However, for some crashes the probability of that happening is so low that it is not even visible in the plot.

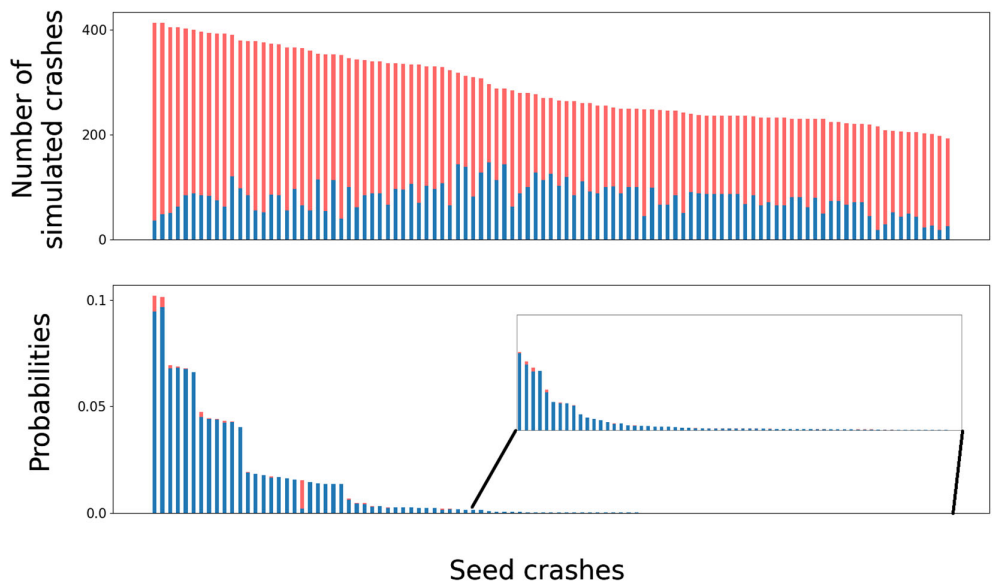

*Figure F1: The top panel shows the number of simulated crashes based on the CBM for each original GIDAS crash (top), sorted in descending order. The bottom panel shows the sum of the products of the probability of the off-road glance and the probability of the deceleration for each individual generated crash, across all the generated crashes for each individual seed crash. Note that all seed crashes result in generated crashes for CBM. The red area (at the top in each bar) represents the crashes that crashed at maximum crash-causation (i.e., the drivers did not brake at all).*

The proportion of crashes with maximum delta-v in the top panel is 72.6%; when prevalence weighting is considered (bottom panel), it is 3.7%.

Figure F2 shows the number of crashes generated in the baseline generation per original (seed) GIDAS crashes for BLOM. Here the white part (columns) in the bottom panel is the 40.8% of the BLOM crashes that are not included. Note the large differences between Figure F1 (CBM) and Figure F2 (BLOM), in both the top and bottom panels. The difference is largest between the (probability-weighted) bottom panels. That is, there is a substantial difference in how the simulated crashes for each individual seed crash contribute to the assessment metrics (i.e., crash frequency and delta-v). It is particularly interesting to note the difference in the proportion of simulated crashes that crash at maximum delta-v: for BLOM (red/upper part of the bars in the lower panel in Figure 8) the proportion is 56.7%, while it for CBM it is 3.7% (bottom panel in Figure 5). Note that this is without adding any crashes that are due to sleeping drivers. These no-reaction crashes are only added for the CBM, resulting in ~14% of crashes with maximum delta-v for BLOM. In general, the differences across the seed crashes are mainly due to difference in the time headway gap (how close the vehicles are to each other) and in how hard the lead vehicle brakes in the seed crashes.

Note that as the individual generated crashes are weighted with the seed crash weight (the inverse of the lower panel in Figures F1 and F2), the crashes generated by the leftmost seed crashes will have a negligible influence on the mean delta-v—much less than those generated by the rightmost ones. This difference holds even if the leftmost seed crashes generate more crashes than the rightmost ones (upper panels of Figures F1 and F2).

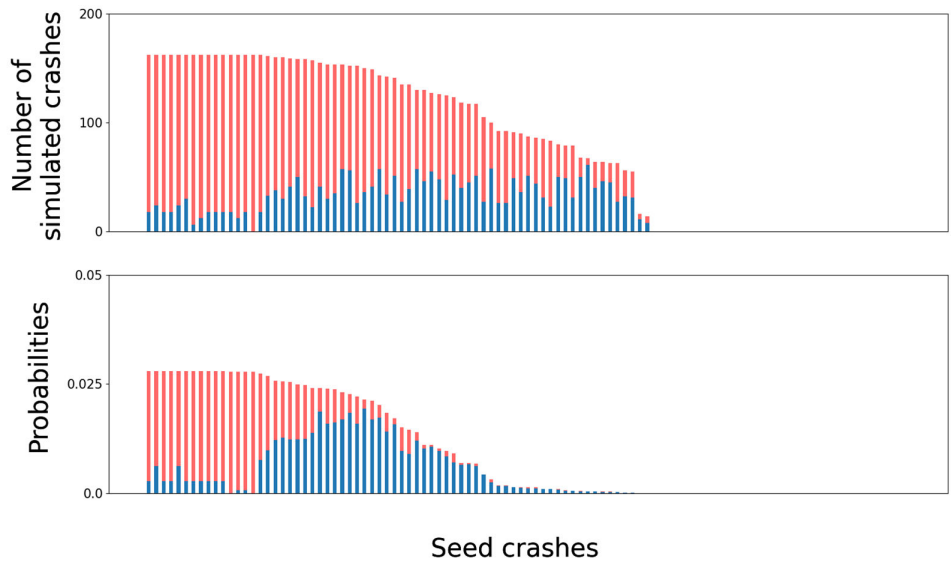

*Figure F2: The top panel shows the number of crashes generated in the baseline, per original (seed) GIDAS crash, for the BLOM. The bottom panel shows the sum of the products of the probability of the off-road glance and the probability of the deceleration for each generated crash, across all the generated crashes for each individual seed crash. Note that the missing value bars for the BLOM (top panel) are the 38 cases that were excluded because the lead vehicle either did not brake or was standing still at the start. Naturally those are also missing from the lower panel.*

Finally, he large difference in the number of crashes per seed crash can have several different reasons, but the most prominent are likely the time gap (how close the vehicles are to each other), and how hard the lead-vehicle brakes in the seed crashes, but also the initial speed has an impact. Consequently, the higher the number of simulated crashes in one case, the 'riskier' the seed crash.

**Appendix G – Sensitivity analysis of the manual data manipulation of the PDO**

A sensitivity analysis was made to assess how sensitive the parameter fitting and simulation outcome (mean delta-v) are to the shape of the data of the (manual added) missing PDO data at the lowest delta-vs. That is, to assess how sensitive the fitted parameters and the delta-v outcome are to the specific choices of values added to the five first bins (i.e., below the mode of the original PDO distribution; the lowest impact speeds). The sensitivity analysis was made with respect to a) the parameters of the fitted exponential function of the PDO, b) the parameters of the resulting transformation function, and c) the outcome delta-v mean (from the simulations). The parameter values for the PDO exponential can be seen in Figure G1, and the parameter values for the transform in Figure G2.

The starting point of the sensitivity analysis was the original manual "best fit" distribution presented in Figure 7. (That is, the height of each of the five bins was iteratively modified to fit the fitted exponential.) The parameter fit outcome can be seen in the top left panels of Figures G1 and G2. A first sensitivity check was to see how large the parameter and delta-v difference would be if a distribution more like a log-normal distribution (manual manipulation of the five lowest five bins to create something log-normal-like). This can be seen in the second panel from the left at the top of Figures G1. The remaining 18 panels in Figures G1 and G2 are random manipulations of the original manual "best fit" exponential (top left). For each bin the original added bin value was changed by adding a (uniformly) random value between -30% to 30% of the bin's value in the original (best) fit. That is, each bin was changed by a random value, and the PDO, and subsequently, the transform, was fitted (as described in the main paper). The minimum and maximum values of parameter c across all cells in Figures G1 and G2 are specifically marked.

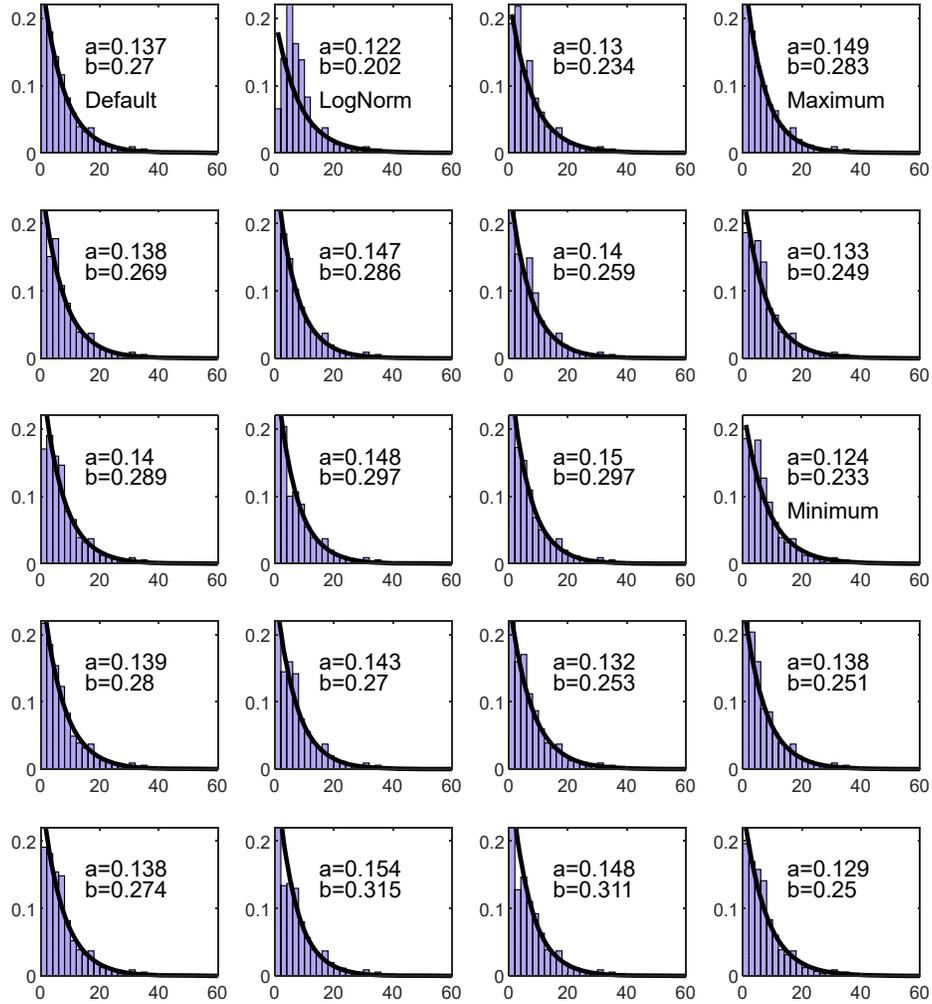

*Figure G1: Parameter fits for the inverse exponential for "all" PDOs (Folksam PDO + added PDOs to create a 70% total PDO in the Folksam data), for the default "best fit", a log normal, and 18 random variants. The minimum and maximum c parameter (in G2) is marked.*

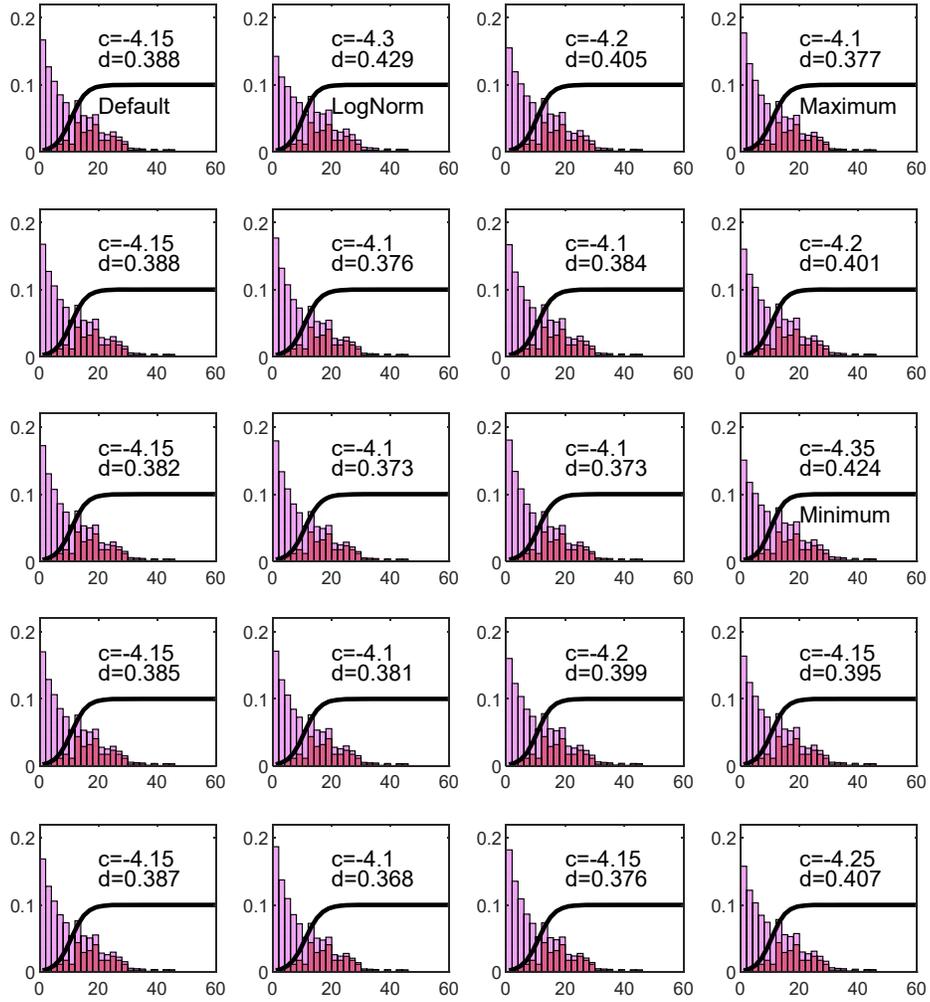

*Figure G2: Parameter fits for the transfer functions associated with the exponential fits in Figure G1. The minimum and maximum c parameter is marked.*

**Appendix H – Distribution summary statistics**

For an enthused reader, Table H1 shows summary statistics used to compare the different models in this work. Below each of the metrics is described in turn.

- Absolute difference from mean – the difference in mean of the compared distributions
- Mean absolute difference – the mean of the difference of each individual impact speed bin between the compared distributions, without taking the probability of each bin into account.
- Weighted mean absolute difference – the mean of the difference of each individual impact speed bin probability weighted using the probabilities of each respective bin.
- Maximum absolute difference – the maximum difference between two bins between the two compared distributions.
- Total variation distance – the normalized L1 distance between two distributions, indicating how much the distributions overlap or diverge:
  $$D_{TV}(p,q) = \frac{1}{2}\sum_{i=1}^{N}|p(x_i) - q(x_i)|,$$
  where N denotes the number of distribution bins, $p(x_i)$ is the weight of the i:th bin in the generated output distribution and $q(x_i)$ is the weight of the corresponding bin in the distribution used for comparison (in our case, the GIDAS data). (Gibbs & Su., 2002; Kullback & Leibler, 1951)
- KL-divergence – the difference in information (entropy) between two distributions, calculated as:
  $$D_{KL}(p,q) = \sum_{i=1}^{N} p(x_i) \log \frac{p(x_i)}{q(x_i)}.$$
  Since the KL-divergence cannot handle non-zero bin weights, half a count was added to each distribution bin and the resulting corrected distributions were then re-normalized. (Gibbs & Su., 2002)
- KS test statistic – the maximum vertical distance between the empirical CDF:s of the compared distributions:
  $$D_{KS}(p,q) = \sup_{i \in \{1,\ldots,N\}} |p(x_i) - q(x_i)| \text{ (Kolmogorov, 1992)}$$

Note that for most of the metrics several weighting steps are needed (i.e., case/propensity weighting, selection bias transform, and the probability that each case would occur, given the off-road glance and maximum deceleration probability). These steps are described in each respective method description in the main paper.

*Table H1: Distribution-comparison statistics where different models are compared with the original delta-v distribution.*

| Metric | GIDAS vs. non-transformed CBM (SHRP2 glances) | GIDAS vs. non-transformed BLOM | GIDAS vs. transformed CBM (SHRP2 glances) | GIDAS vs. transformed BLOM |
|---|---|---|---|---|
| Mean (GIDAS original= 17.84 km/h). [km/h] | 13.63 | 19.50 | 17.50 | 22.55 |
| Absolute difference from (full) mean. [km/h] | 4.25 | 1.66 | 0.34 | 4.70 |
| Mean (full) absolute difference. [km/h] | 0.0111 | 0.00712 | 0.00631 | 0.00844 |
| Weighted (full) mean absolute difference. [km/h] | 0.0329 | 0.0215 | 0.0236 | 0.0250 |
| Max absolute difference. [km/h] | 0.0904 | 0.0734 | 0.0560 | 0.0653 |
| Total variation distance | 0.332 | 0.214 | 0.189 | 0.253 |
| KL-divergence | 0.484 | 0.222 | 0.178 | 0.291 |
| KS-distance | 0.318 | 0.122 | 0.055 | 0.244 |

**Appendix I – Details limitation related to other crash-causation factors**

The high-level statistics of factors, not directly related to the GIDAS PCM, were studied from the included cases in the GIDAS database in order to consider whether the developed model would be expected to cover them. The two main categories, 'environment' and 'driver', are discussed below. Just because these factors are coded in GIDAS does not mean they are always the main cause of the crash, but they are likely to be contributing factors. One must also remember that the lack of reports on these factors does not necessarily mean that they are not important in reality.

Parameters identified in GIDAS as environment-related factors include rain and snow and their effect on braking performance, as well as rain, snow, fog, and darkness and their effect on glance behavior and visibility. The factors affecting braking performance is in this work partially captured by the use of a distribution of maximum driver decelerations. However, the CBM does not, for now, consider visibility factors.

Driver-related factors include, for example, medical conditions such as heart attack, seizures, cramps, and fatigue. The 10% added to the model to represent sleepy drivers will cover these factors to some extent, but further investigations should be made into their prevalence. Note that if it is found that these factors play a larger role (than what is captured in the 10% no-reaction crashes) it is possible to adjust the percentage no-reaction crashes. Further, in-depth crash investigations have shown that factors such as the view of the target being obstructed by the sun visor, glare from the sun, or other traffic participants also may contribute to crashes. However, since the model assumes perfect visibility when there is no off-road glance, these factors are not explicitly captured in the CBM. However, also if future research indicates that such factors are a substantial part of crashes, the proportion of no-reaction crashes may be adjusted also to cover these factors.

There are also underlying driver parameters such as age, general medical conditions, and medications—as well as the current state of the driver (including the use of alcohol or other drugs). The variability of the input data (glance behavior data and maximum deceleration, as well as other parts of the CBM) likely capture some of these, but far from all. Future studies may go into detail on these factors to investigate if it is possible (and there is a need) to consider them in the crash-causation models (e.g., introducing minor changes in the response model).

Further driver-related factors are related to the driver's ability to judge the absolute and relative speeds and distances in traffic, as well as mismatches between driver expectations and what actually happen on the road (e.g., if a lead-vehicle driver is suddenly aborting a overtaking). Also these factors are partially captured in the glance behavior and maximum deceleration data, but further research is needed, especially on how to model expectation mismatches (see Bianchi Piccinini et al., 2020, for one example of expectation mismatch in an ADS setting).

**Appendix J – Analysis of the differences and similarities of the BLOM and CBM delta-v distributions**

We further analyzed the difference between BLOM and CBM delta-v distributions in detail, compared to the seed delta-vs. That is, we compared the delta-v distributions between BLOM and CBM for the individual seed crashes (Figure J1 shows a few examples of this type of comparison), studied the time-series data, and watched animations of the pre-crash kinematics. First, we must remind the reader that all the generated crashes for a seed crash jointly contribute to 1/N of the overall delta-v distribution (e.g., Figure 7 in the main manuscript), where N is the number of seed crashes (103 for CBM and 68 for BLOM). Furthermore, each simulation has its inherent probability of generating a crash, coming from the joint probability of an off-road glance and vehicle deceleration. Thus, for seed crashes that in general are easy to avoid for the model (either CBM and BLOM), only a few crashes are generated. Each of these 'easy-to-avoid' generated crashes will contribute much more, relatively, to the overall delta-v distribution than any of the numerous generated crashes that are not as easily avoided. Also note that, when only a few crashes are generated for a seed, they are typically clustered in a relatively narrow range of delta-vs, making their contribution to the overall delta-v distribution particularly strong (compared to more crashes spread over a larger range, which would then contribute less at each delta-v).

It turns out that not only do the 'easy-to-avoid' crashes have more weight when pooled in the overall delta-v distribution, but the kinematics of these crashes are similar for the two models. One example of an 'easy-to-avoid' crash for BLOM is: the following vehicle is at a long THW when the lead-vehicle starts to brake (brake light onset), and the lead vehicle has come to a stop before collision. The 'easy-to-avoid' cases for the CBM model are cases with a low relative speed in combination with a proportionally large THW at the time of $\tau^{-1}$=0.2. That is, the probability of crashing in both the BLOM and the CBM example is very low, since it would require a long reaction time (or off-road glance) and/or low deceleration. As the kinematics for the 'easy-to-avoid' crashes are similar for both CBM and BLOM, they contribute similarly to the delta-v distribution. However, only a subset of the seed crashes are 'easy-to-avoid' (and thus similar). The analysis of the individual delta-v distributions showed that many of the BLOM distributions are 'narrow' and have substantially higher delta-vs than the delta-v of the seed – this is confirming what can be seen in the percentile plots (Figure 9). Different from BLOM, CBM distributions are wider, and in general the average delta-v per case is closer to the seed delta-v.

An example of such an analysis, but with only four example seeds, are shown in Figure J1. Panels a and b in Figure J1 show similar delta-v distributions for both BLOM and CBM. However, in panel a, CBM also has a distinct peak (at delta-v just above 40 km/h) corresponding to the sleepy driver reactions of the CBM model. The actual delta-v for the seed crash coincides with that peak, indicating that this is a case where the actual driver was unresponsive. Panel c and d are examples of seed crashes where CBM and BLOM differ substantially and only one of the models match the actual delta-v of that seed (but note that on an individual case level the seed crash can really be 'anywhere' as it just is one instance of many possible crashes). In both cases, BLOM results in a narrower distribution with emphasis on higher delta-vs than CBM. Since the actual delta-v happens to be high for the case in panel c, due to an unresponsive driver, BLOM performs well on this specific case (but, again, that does not mean that it is "better" as a model). Note that to make the distributions easier to read, a smoothing kernel has been used; this has however also smoothened out the sleepy driver peaks (highest delta-v values in the figure), which in reality are single delta-v values. The peaks at the highest delta-vs in the CBM data are the 10% sleepy driver contribution.

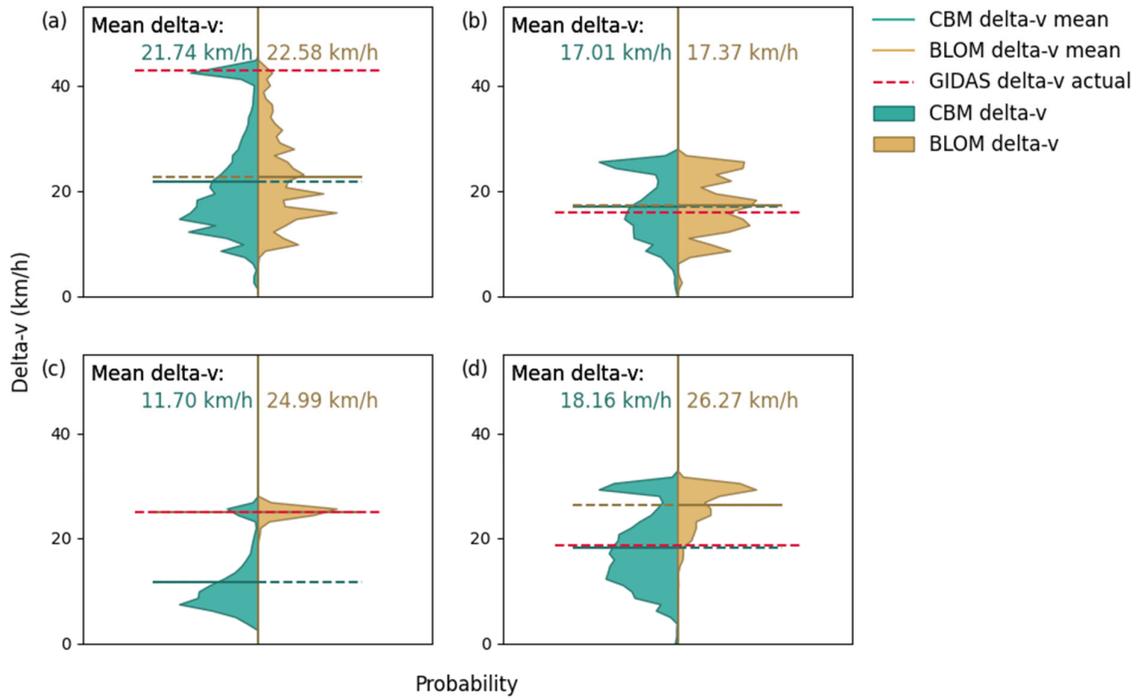

*Figure J1: Illustrations of the delta-v distribution of the generated crashes for CBM and BLOM, respectively, for four example seed crashes.*

**Appendix K – Description of the percentile validation**

The percentile histogram (distribution) assessment used in this work can be seen as the inverse of random number generation from a specific distribution. That is, one way of creating a random number is to select a percentile with a uniform probability 0–100%, then use the value on the x-axis as the random numbers (see Figures K1 and K2). What we do in this work is that we have values on the x-axis (delta-v here) instead and get the percentiles values for those. If the set of delta-v values you have then produces a uniform distribution of percentiles, that is an indication that the values are random values from that distribution (i.e., the underlying data generation process for the data; see Figure K3). Consequently, if you had the underlying distribution to start with and would randomly have sampled with a uniform percentile distribution, you would get the shape of the delta-vs that you checked for uniformity in the percentile distribution for (that is, you would recreate the same shape). In this work that would be equivalent to creating a cumulative distribution function of the original GIDAS seeds, split it into percentiles, and then sampling uniformly from the percentiles – with enough samples, the result would be the same distribution as the seeds.

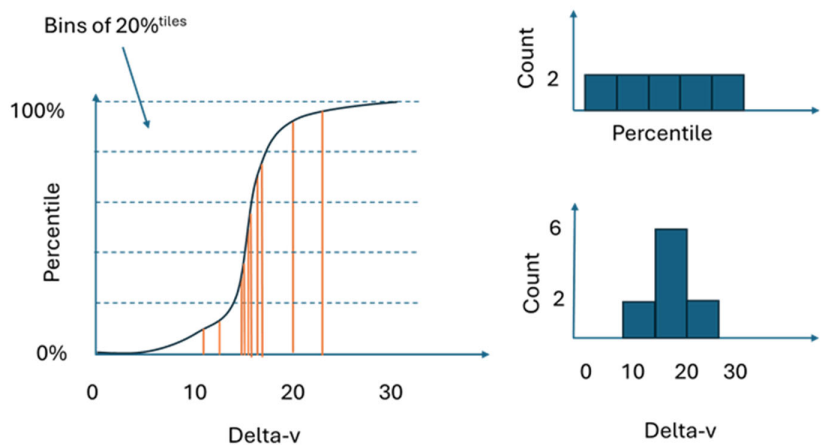

*Figure K1: An illustration of the process of random sampling from the percentiles in a delta-v distribution. Here 20-percentile bins are used, and two values are 'sampled' from each bin. As a result, the percentile histogram is uniform (two counts per bin), while the delta-v distribution is whatever it is. If we instead had the individual delta-v values and wanted to check if the data really came from a specific distribution (the black curve), we would create the percentile histogram from the delta-v values. In this case it would show that the data came from that distribution.*

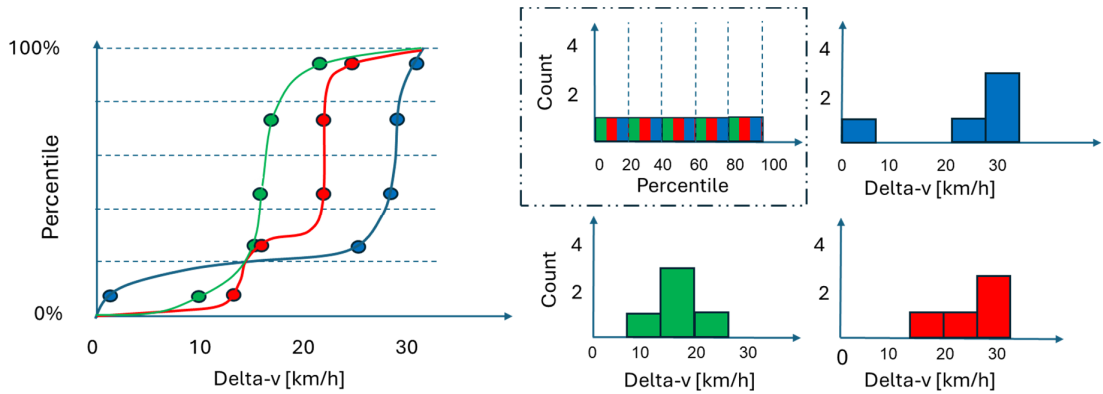

*Figure K2: Conceptual illustration of the data generation process for a 'random number generator' from percentiles for three different distributions. That is, 'sampling' one value per percentile bin (naturally) produces different delta-v distributions. Note that here the y-values are the same for all three distributions (as opposed to Figure K3, where the x-values are the same), except when they are overlapping, where they are then shown partially overlapping.*

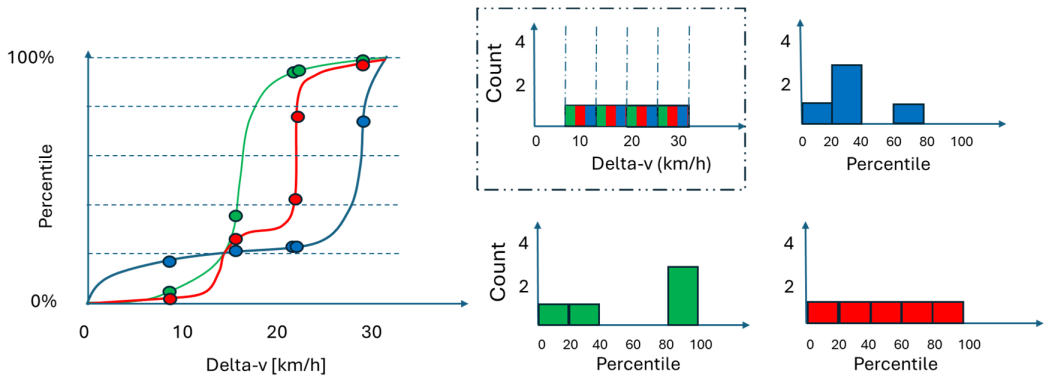

*Figure K3: This figure conceptually illustrates checking if the given delta-v values (x-axis in left panel) come from any of the three given distributions. Note that the x-values are the same here for all three distributions (unlike Figure K2). Here the red distribution is the most likely, as it produces a uniform distribution. As there are only five datapoints, the uniformity could also happen by chance in this case. However, as more datapoints are added, if the transform from delta-v values to percentiles (given a specific distribution) produces a uniform distribution, it becomes more and more likely that the data is from that distribution (which is equivalent to saying it comes from the same underlying data generation process).*

We also made a final percentile analysis to investigate the sensitivity of the choice of the number of bins used in the distribution (normalized histogram). Remember that there is a total of 103 CBM seed crashes and 68 BLOM crashes, which is the total count (turned into frequency) in the distribution. That means that if the distributions are uniform, the average "bin height" is 103 (for CBM) and 68 (for BLOM) divided by the number of bins (and normalized). As a result, the higher the number of bins, the more sensitive the distribution is to the bin edges and randomness in the sample. The sensitivity analysis reveals that the CBM distributions have a consistent shape, regardless of the bin size. We observe higher proportions of seed delta-vs at high percentiles (60–90th), and lower proportions of seed delta-vs at lower percentiles (<40th percentile). In contrast, the shape of the BLOM distributions do not show the same consistency. Instead, the bin corresponding to the lowest percentile gets more prominent as the number of bins increases. The dominance of very small seed delta-v percentiles in the distributions generated by BLOM indicates that most seed delta-vs can be found at the left tail.

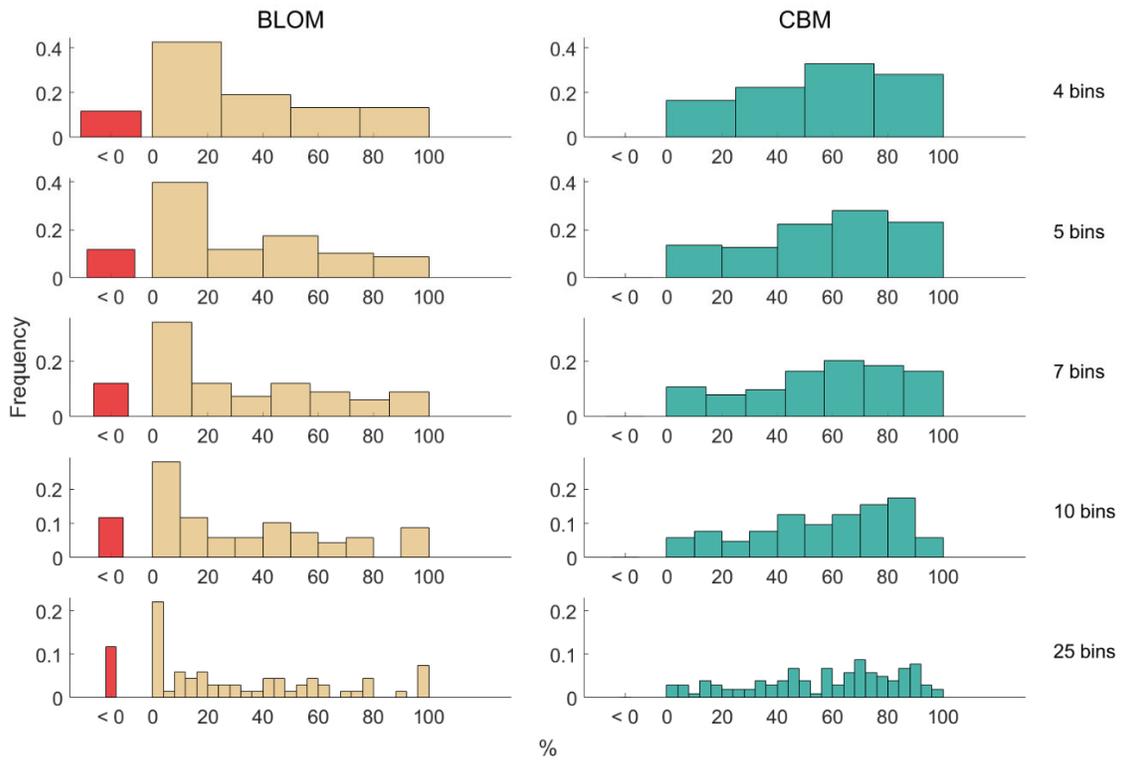

*Figure K4: Shows how the number of bins in the empirical distribution impacts the overall distribution shape.*